\documentclass{jfm}

\usepackage{epstopdf, epsfig}
\usepackage{graphicx,footmisc}
\usepackage{subfig,color}
\usepackage{natbib,url}
\usepackage{lineno,amsmath}
\usepackage{float}
\usepackage{xcolor}

\newcommand*{\be}{\begin{equation}}
\newcommand*{\ee}{\end{equation}}

\def\begineq{\begin{equation}}
\def\endeq{\end{equation}}

\def\begineqn{\begin{equation*}}
\def\endeqn{\end{equation*}}

\def\beginar{\begin{eqnarray}}
\def\endar{\end{eqnarray}}
\def\beginarn{\begin{eqnarray*}}
\def\endarn{\end{eqnarray*}}

\def\lb{\left ( }
\def\rb{\right ) }

\def\ub{\mathbf{u}}

\def\dst{{\partial_t}}

\def\dsz{{\partial_z}}

\def\hz{{\bf\widehat z}}

\shorttitle{Quasi-static magnetoconvection}
\shortauthor{M. Yan et al.}

\title{Heat transfer and flow regimes in quasi-static magnetoconvection with a vertical magnetic field}%

\author{Ming Yan$^1$, 
Michael A. Calkins$^1$,
Stefano Maffei$^1$,
Keith Julien$^2$, 
Steven M. Tobias$^3$, and
Philippe Marti$^4$}

\affiliation{
$^1$Department of Physics, University of Colorado, Boulder, CO  80309, USA \\
$^2$Department of Applied Mathematics, University of Colorado, Boulder, CO  80309, USA \\
$^3$Department of Applied Mathematics, University of Leeds, Leeds, UK LS2 9JT \\
$^4$Department of Earth Science, ETH Zurich}
\begin{document}

\maketitle

\begin{abstract}
Numerical simulations of quasi-static magnetoconvection with a vertical magnetic field are carried out up to a Chandrasekhar number of $Q=10^8$ over a broad range of Rayleigh numbers $Ra$. Three magnetoconvection regimes are identified: two of the regimes are magnetically-constrained in the sense that a leading-order balance exists between the Lorentz and buoyancy forces, whereas the third regime is characterized by unbalanced dynamics that is similar to non-magnetic convection. Each regime is distinguished by flow morphology, momentum and heat equation balances, and heat transport behavior. One of the magnetically-constrained regimes appears to represent an `ultimate' magnetoconvection regime in the dual limit of asymptotically-large buoyancy forcing and magnetic field strength; this regime is characterized by an interconnected network of anisotropic, spatially-localized fluid columns
aligned with the direction of the imposed magnetic field that remain quasi-laminar despite
having large flow speeds.
As for non-magnetic convection, heat transport is controlled primarily by the thermal boundary layer. Empirically, the scaling of the heat transport and flow speeds with $Ra$  appear to be independent of the thermal Prandtl number within the magnetically-constrained, high-$Q$ regimes.

\end{abstract}

%\begin{keywords}
%\end{keywords}

\section{Introduction}

Convective heat transfer is a fundamental process that controls the thermal evolution of planets and stars \citep{mM05,cJ11b}. In these natural systems the fluid is strongly forced, and thought to be in a turbulent state.
Magnetic fields generated by the motion of electrically conducting fluids permeate many of these systems, and can have a significant influence on the dynamics via electromagnetic forces. Understanding such dynamics is crucial for determining how planets and stars evolve thermally over their lifetimes. However, the detailed role of strong magnetic fields in modifying the heat transport and dynamics remains poorly understood when the buoyancy forcing becomes large.

Rayleigh-B\'enard convection is a canonical model for theoretical and numerical studies of buoyancy-driven flow that consists of a fluid layer contained between plane, parallel boundaries separated by a vertical distance $H$. A constant gravity vector  ${\boldsymbol g} = -g \hz$ points vertically downward ($\hz$ is the vertical unit vector), and a constant temperature difference  $\Delta \mathcal{T} = \mathcal{T}_{bottom} - \mathcal{T}_{top} > 0$, is maintained to drive convection. For a Boussinesq fluid with thermal expansion coefficient $\alpha$, kinematic viscosity $\nu$, and thermal diffusivity $\kappa$, convective motions are controlled by the Rayleigh number ($Ra$) and thermal Prandtl number ($Pr$),
\be
Ra = \frac{ g \alpha \Delta \mathcal{T} H^3}{\nu \kappa}, \quad Pr = \frac{\nu}{\kappa} .
\ee

As the Rayleigh number becomes large,  unconstrained convection is known to undergo a transition to turbulence, as characterized by a broad range of spatiotemporal scales \citep[e.g.][]{gA09,lohse2010small,chilla2012new}.

When an externally-imposed, vertical magnetic field $\bold{B}_0 = \mathcal{B} \, \hz$ permeates the fluid layer, the convective dynamics also depends on the Chandrasekhar number ($Q$) and magnetic Prandtl number ($Pm$) defined as
\be
Q = \frac{\mathcal{B}^2 H^2}{\rho \nu \mu \eta}, \quad Pm = \frac{\nu}{\eta} ,
\ee
where $\mathcal{B}=|\bold{B}_0|$, $\rho$ is the fluid density, $\mu$ is the vacuum permeability, and $\eta$ is the magnetic diffusivity. The relative sizes of the thermal and magnetic Prandtl numbers control the time-dependence of the onset of convection; for fluids characterized by $Pr \ge Pm$, the onset of convection is steady, whereas oscillatory convection occurs when $Pr < Pm$ \citep{sC61}. The former relationship is relevant to liquid metals, including both planetary interiors \citep{mF12,mP13} and laboratory experiments \citep[e.g.][]{sC00,jmA01,uB01,nG07b,tY10,eK13,eK15,tV18}. Stellar interiors composed of plasmas are typically characterized by $Pr < Pm$ \citep[e.g.][]{mO03}, so oscillatory convection is likely important in this context. In the present work we consider the magnetohydrodynamic quasi-static limit only; the induced magnetic field is asymptotically-small relative to the imposed magnetic field and, as a result, the onset of convection is always steady \citep{sC61}.

For an asymptotically-strong vertical magnetic field, $Q\rightarrow \infty$, it can be shown that, within a layer of infinite horizontal extent, the onset of (steady) convection is characterized by critical Rayleigh number $Ra_c \rightarrow O( \pi^2 Q)$ and critical horizontal wavenumber $k_c \rightarrow O(\frac{1}{2} (\pi^4 Q)^{1/6})$ \citep{sC61,pM99}. Thus, the presence of a vertical magnetic field acts to stabilize convection, and leads to anisotropic motions.

Strongly-forced nonlinear magnetoconvection (MC) with $Q \gg 1$ remains poorly understood, despite its relevance for natural systems. For instance, estimates for the magnetic field strength in the Earth's outer core range up to $Q \approx 10^{15}$ \citep[e.g.][]{nG10}. In contrast, laboratory experiments and numerical simulations have been limited to $Q \lesssim O(10^6)$  \citep{sC00,jmA01,uB01,lT98,fC03,tZ16,xY18,wL18}. Both laboratory experiments \citep{jmA01} and numerical simulations \citep{xY18} have found a non-dimensional heat transport scaling of $Nu \sim ( Ra/ Q)^{1/2}$ for $Q \lesssim 10^3$, where $Nu$ is the Nusselt number. In contrast, the experimental study of \cite{uB01} suggests a $Nu \sim ( Ra/ Q)^{2/3}$ scaling for $Q \rightarrow \infty$, though accessible values of the Chandrasekhar number were limited to $Q \lesssim 10^3$. The experiments of \cite{sC00} reached up to $Q=3.93 \times 10^6$ and covered a broad range of supercritical Rayleigh numbers in which three heat transport regimes were observed (in order of increasing $Ra$): (1) a $Nu \sim (Ra/Q)$ regime (their regime I); (2) an intermediate regime (their regime III) in which the heat transport law varies continuously with increasing $Ra$; and (3) a third regime (their regime II) in which $Nu \sim Ra^{0.43} Q^{-0.25}$.

Scaling predictions for the heat transport in MC have used Malkus's (\citeyear{wM54}) concept of a marginally stable thermal boundary layer \citep{jB91b}, and energetic arguments \citep[using the approach introduced by][]{sG00} relying on a predominance of ohmic dissipation over viscous dissipation \citep{sB06}. These two assumptions both lead to a $Nu \sim (Ra/Q)$ heat transport scaling law as $Q \rightarrow \infty$. Interestingly, this scaling law is independent of the height of the domain $H$, and independent of all diffusion coefficients except the magnetic diffusivity. This latter property suggests that the heat transport  scaling behavior is independent of $Pr$ as $Q \rightarrow \infty$.

In the present work we carry out direct numerical simulations of quasi-static MC in the plane layer  geometry with magnetic field strengths up to $Q = 10^8$. We find three unique MC regimes that can be distinguished by flow characteristics, force and heat equation balances, and heat transport ($Nu$) scalings. The first regime is reminiscent of linear convection, with cellular flow structures and a heat transport that increases rapidly but cannot be characterized by a single power-law scaling. The second regime is characterized by localized, quasi-laminar convection `columns' that align with the imposed magnetic field and shows a $Nu \sim (Ra/Q)^{\gamma}$ scaling, but with a value of $\gamma$ that increases toward unity with increasing $Q$. Thus, our findings indicate that the previously observed $Nu \sim (Ra/Q)^{1/2}$ and $Nu \sim (Ra/Q)^{2/3}$ scalings are \textit{transitional} and limited to relatively small values of $Q$. A third MC regime is observed that is similar to non-magnetic convection in both flow structure and heat transport behavior; here the flow is observed to become broadband in structure. Our results suggest that quasi-static MC does not become turbulent provided the Lorentz force remains dominant -- we refer to such states as `magnetically-constrained'. Thus, two magnetically-constrained regimes are identified, whilst the third regime might be characterized as `magnetically-influenced'.

\section{Methods}

We use the quasi-static magnetohydrodynamic approximation that is valid when the magnetic Reynolds number  $Rm = Pm Re \rightarrow 0$, where the hydrodynamic Reynolds number is defined as $Re = U L/\nu$ ($U$ is a typical flow speed, $L$ is a typical flow lengthscale) \citep[e.g.][]{hM70b}.  In particular, the magnitude of the induced magnetic field ($\bold{b}$) is smaller than the imposed field ($\bold{B_0}$)  by $O(Rm)$, thus $\bold{b} \sim O(Rm \bold{B_0})$; this model has been used by many previous investigations \citep[e.g.][]{tZ16,xY18,wL18}. Using this limit, the non-dimensional governing equations are given by
\be
\partial_t  \bold{u} = \underbrace{\nabla^2\bold{u}}_{F_v} -\underbrace{\bold{u}\cdot\bold{\nabla}\bold{u}}_{F_a} +\underbrace{Q \partial_z \bold{b} }_{F_l}+ \underbrace{\frac{Ra}{Pr}T^\prime\bold{\hat{z}}}_{F_b}   -\underbrace{ \nabla\Pi}_{F_p}\\\\ ,
\label{eq7}
\ee
\be
0= \nabla^2\bold{b} +\partial_z \bold{u} ,
\label{eq8}
\ee
\be
\left(\partial_t - \frac{1}{Pr}\nabla^2\right)T^\prime= - \bold{u}\cdot\nabla T^\prime-u_z\partial_z\overline{T}+\partial_z\ \overline{\lb u_z T' \rb}, \\\\
\label{eq9}
\ee
\be
\left(\partial_t - \frac{1}{Pr}\partial_z ^2\right)\overline{T}= -\partial_z\ \overline{\lb u_z T' \rb}, \\\\
\label{eq10}
\ee
\be
\nabla\cdot\bold{u} = 0,\quad
\nabla\cdot\bold{B} = 0,\\\\
\ee
\be
\bold{B}=\hz+\bold{b},\quad
T=\overline{T}+T^\prime,
\ee
where $\bold{u}$ is the velocity field,  
$\bold{b}$ is the induced magnetic field, $T$ is the temperature, $\overline{T}$ is the horizontally-averaged (mean) temperature (where $\overline{(\cdot)}$ denotes a horizontal average), $T^\prime$ is the fluctuating temperature and $\Pi$ is the reduced pressure.  Each of the forces present in the momentum equation (\ref{eq7}) have been identified by the symbols below them for future reference in our results. The  viscous force, advection, Lorentz force, buoyancy force and  pressure gradient force are given by $F_v$, $F_a$, $F_l$, $F_b$ and $F_p$, respectively. The horizontal and vertical components of inertia are denoted by $\partial_t u_h$ and $\partial_t u_z$, respectively, where $u_h= \sqrt{{u_x}^2+ {u_y}^2}$, $u_x$  and $u_y$ are the horizontal velocity components, and $u_z$ is the vertical velocity component.
The equations have been non-dimensionalized by the domain-scale viscous diffusion time $H^2/\nu$, imposed magnetic field magnitude $\mathcal{B}$ and temperature difference $\Delta \mathcal{T}$. 
The boundary conditions are stress-free, constant temperature, and electrically insulating.  

The equations are solved using a standard toroidal-poloidal decomposition of the velocity and magnetic field such that the solenoidal conditions are satisfied exactly \citep[e.g.][]{cJ00b}. A  pseudo-spectral code  is used for simulating the above equations with Fourier series in the horizontal dimensions and Chebyshev polynomials in the vertical dimension \citep{pM16}. Numerical resolutions using up to $1536 \times 1536 \times 192$ physical-space grid points are used to ensure that the flow is well-resolved; these resolutions allow for at least 8 vertical grid points within the thermal boundary layer. The non-linear terms are de-aliased with the standard 2/3-rule. The equations are discretized in time with a third-order implicit-explicit Runge-Kutta scheme \citep{pS91}. 
In most of our simulations we use a Prandtl number of $Pr=1$; however, additional simulations with $Pr=0.025$, relevant to liquid metals, suggest that our findings are insensitive to $Pr$.

The horizontal dimensions of the system are scaled by the critical horizontal wavelength, $\lambda_c$. The Rayleigh number corresponding to the marginal stability of horizontal wavenumber $k$ is given by \citep{sC61}
\be
Ra_{m}=\frac{\pi^2+k^2}{k^2}[(\pi^2+k^2)^2+\pi^2Q]  .
\ee
The critical Rayleigh number $Ra_c$ is the minimum value of $Ra_m$ for a given value of $Q$, and is found by minimizing the above expression for all $k$ to find the critical wavenumber $k_c$ that satisfies the expression 
\be
2{k_c}^6+3\pi^2{k_c}^4=\pi^6 + \pi^4Q ,
\ee
where $\lambda_c = 2 \pi/k_c$. For the majority of the simulations we use a domain with non-dimensional size $10\lambda _c\times 10\lambda _c \times 1$. However, as the Rayleigh number is increased, the horizontal dimensions of the system can be reduced while still providing accurate flow statistics. Horizontal dimensions of $5\lambda _c\times 5\lambda _c$ are used for our most extreme cases. Tests with different horizontal dimensions were used to ensure that computed statistics showed convergence.

Amongst the output quantities we analyzed  the Nusselt number, $Nu$, and the Reynolds number, $Re$. 
The Nusselt number measures the efficiency of convective heat transfer in our simulations and is defined by 
\be
Nu = 1 + Pr \langle{ u_z T'}\rangle ,.eps
\ee
where  $T' = T - \overline{T}$, and $\langle \cdot \rangle$ denotes a volumetric and time average.
The Reynolds number measures the typical flow speeds and, with our particular non-dimensionalization of the governing equations, is defined by
\be
Re = \langle{ {u_x}^2+{u_y}^2+{u_z}^2}\rangle^{1/2} .
\ee

Details of the numerical simulations are provided in the Appendix.

\section{Results}

\subsection{Flow regime characterization}

\begin{figure*}

   \subfloat[]{
      \includegraphics[width=0.48\textwidth]{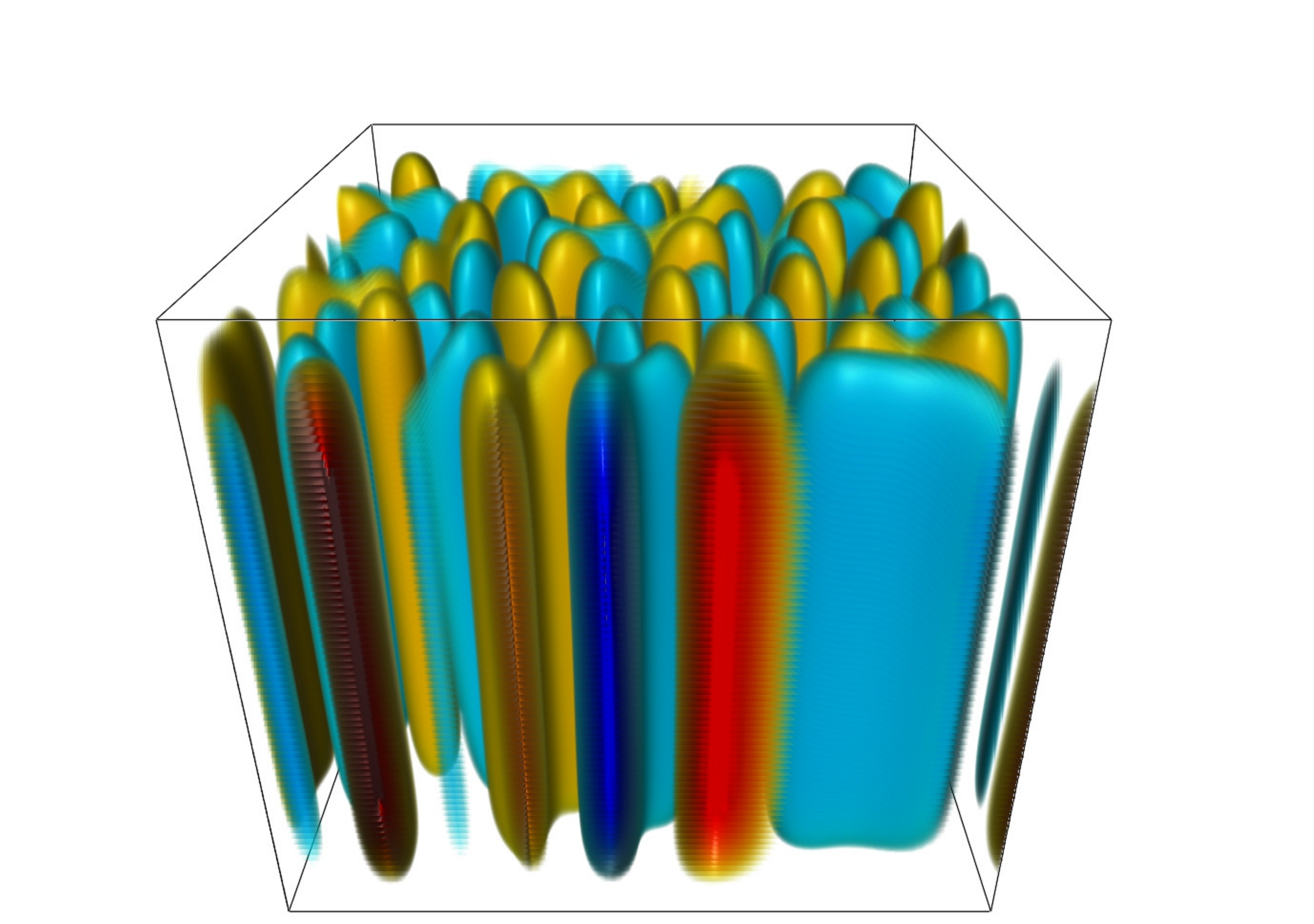} }
         \subfloat[]{
      \includegraphics[width=0.48\textwidth]{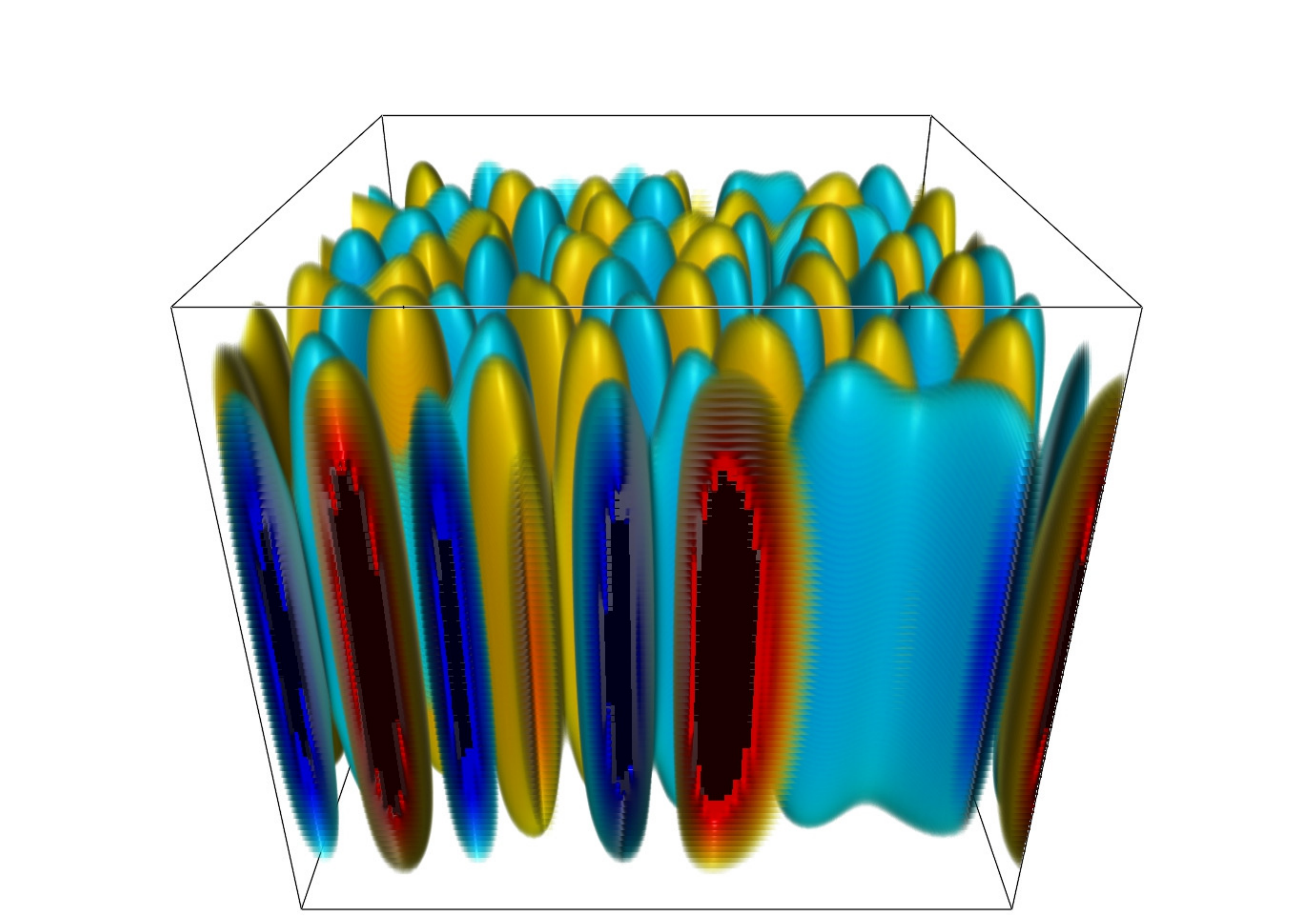} } \\      
      
          \subfloat[]{
      \includegraphics[width=0.48\textwidth]{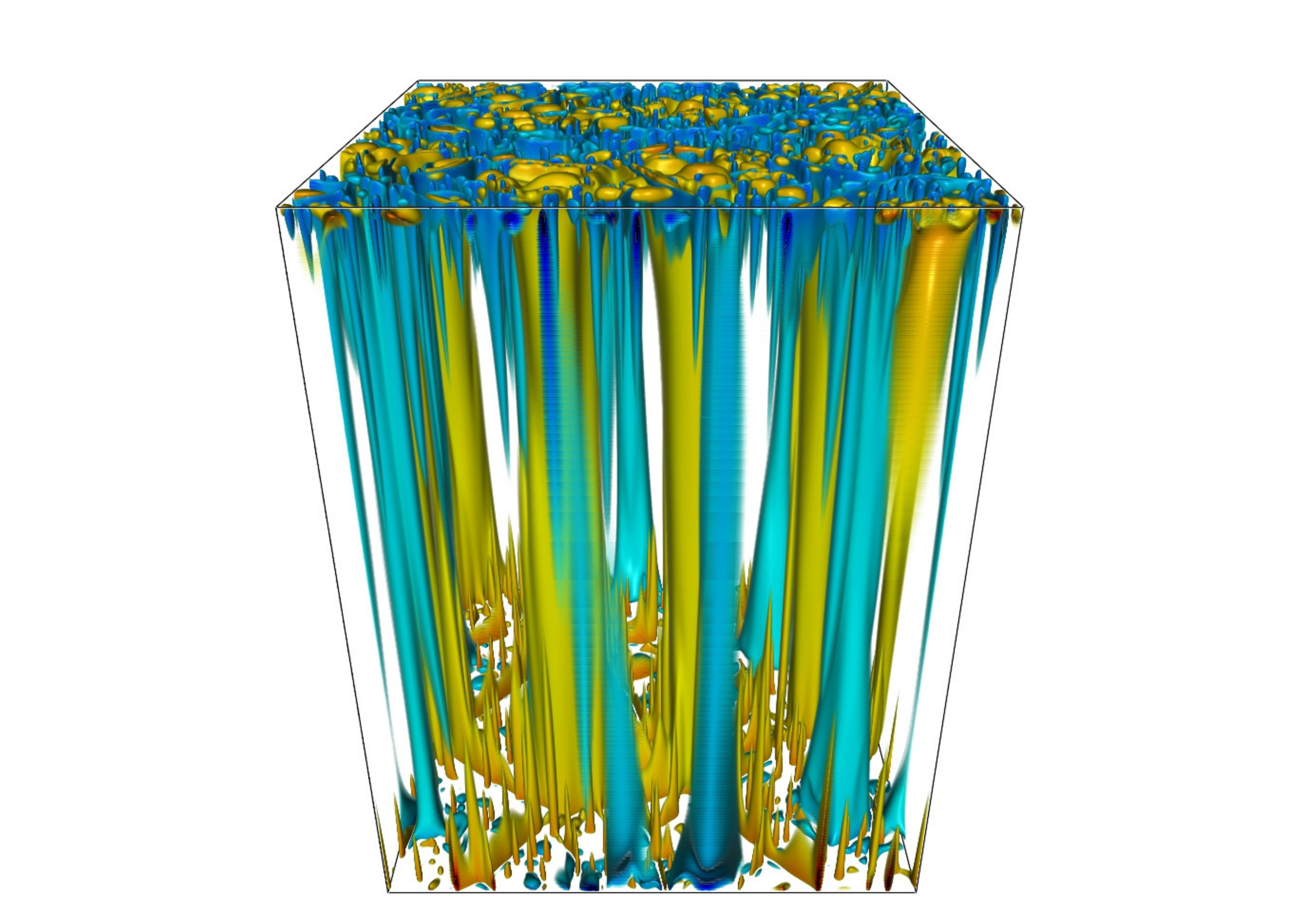}}        
            \subfloat[]{
      \includegraphics[width=0.48\textwidth]{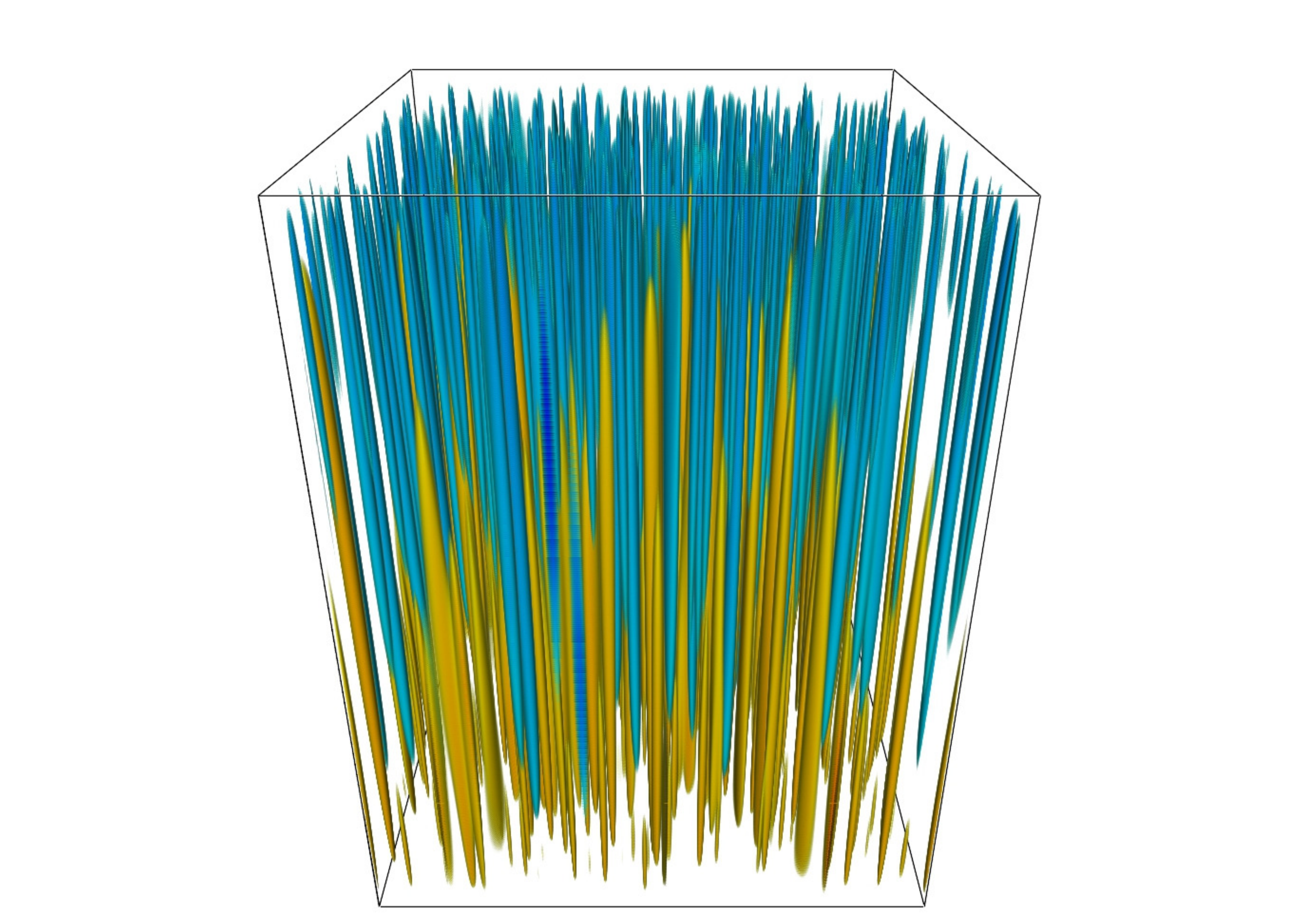} } \\

         \subfloat[]{
      \includegraphics[width=0.48\textwidth]{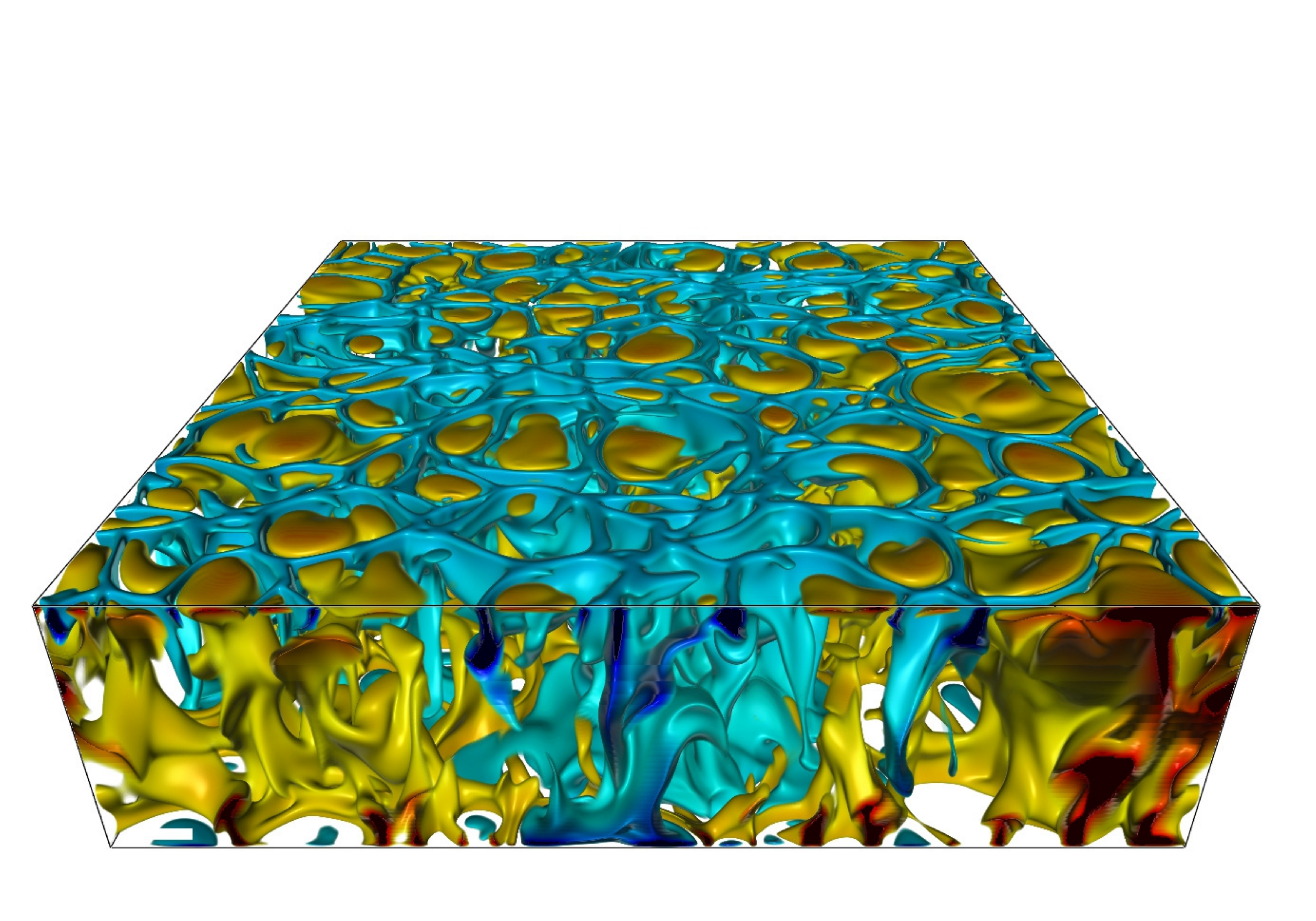} }  
          \subfloat[]{
      \includegraphics[width=0.48\textwidth]{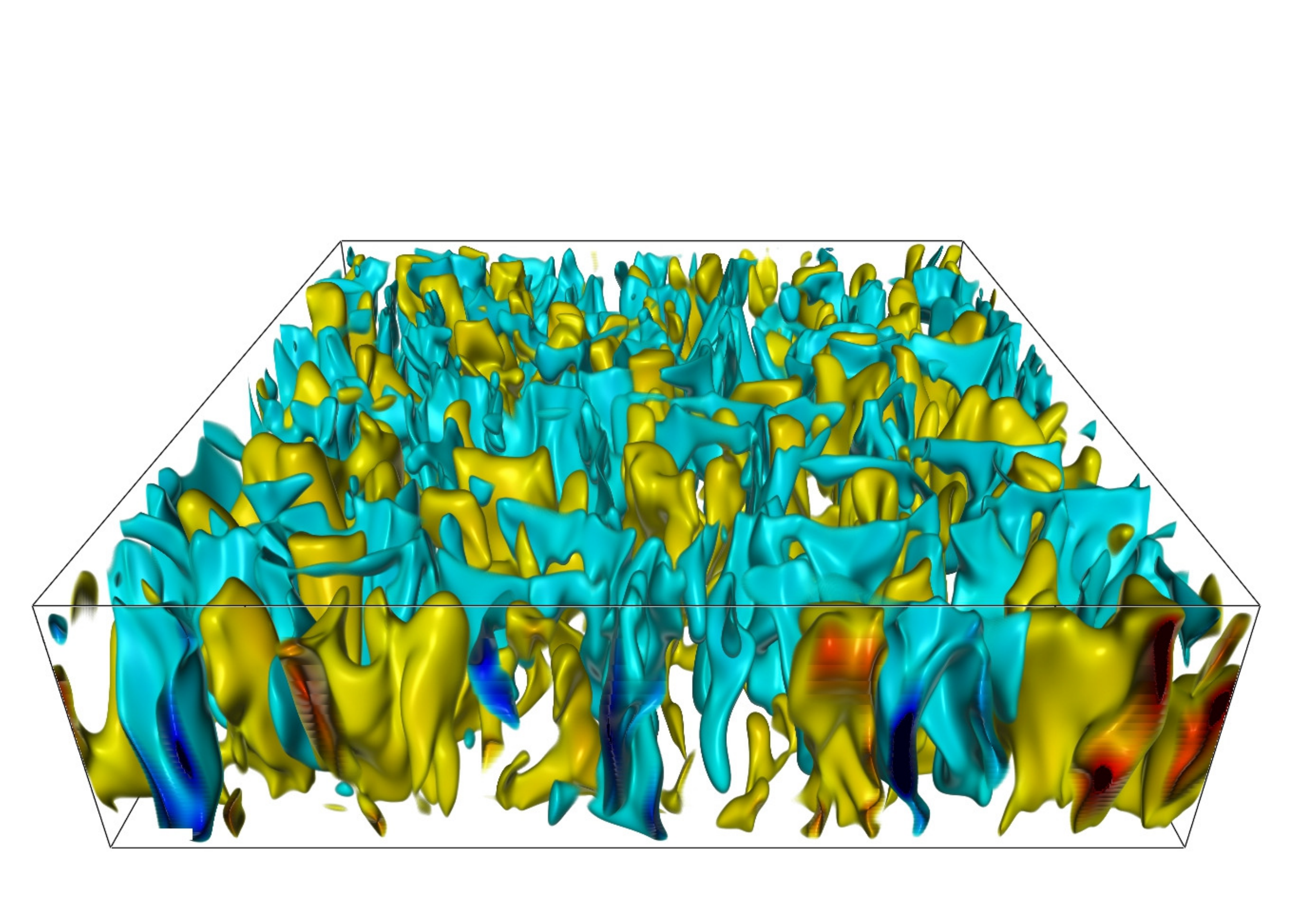}}

\caption{ Volumetric renderings of the three magnetoconvection  regimes identified in the present  study.  Isosurfaces of fluctuating temperature are shown in the first column [(a), (c), (e)]; isosurfaces of the vertical velocity are shown in the second column [(b), (d), (f)]. (a), (b) The cellular regime ($Q=10^7$, $Ra=1.3\times 10^{8}$); 
(c), (d) the columnar regime ($Q=10^8$, $Ra=4\times 10^{10}$); 
(e), (f) the turbulent regime ($Q=10^4$, $Ra=2\times 10^{7}$). 
For all three cases, the Prandtl number is fixed at $Pr=1$, and an aspect ratio of $5\lambda _c\times 5\lambda _c\times 1$ is shown, where $\lambda_c$ is the critical wavelength (see main text for details). 
\label{F:flow_regimes}}
\end{figure*}

Three primary dynamical regimes of MC are found, which we refer to as the cellular, columnar and turbulent regimes. Each regime is illustrated in Fig.~\ref{F:flow_regimes}, where each panel shows a simulation domain with aspect ratio $5\lambda _c\times 5\lambda _c\times 1$, where $\lambda _c\approx 0.22$, $0.15$ and $0.73$, respectively, for the three different cases. Only the first two regimes are considered to be magnetically-constrained in the sense that the Lorentz force plays a leading-order role in the dynamics. Each regime can be uniquely identified by: (1) the scaling of the heat transport and flow speeds with buoyancy forcing; (2) the physical structure and spectral characteristics of the flows; and (3) the relative sizes of each term in the governing equations.

\begin{figure*}
  \begin{center}
      \subfloat[]{
      \includegraphics[height=5cm]{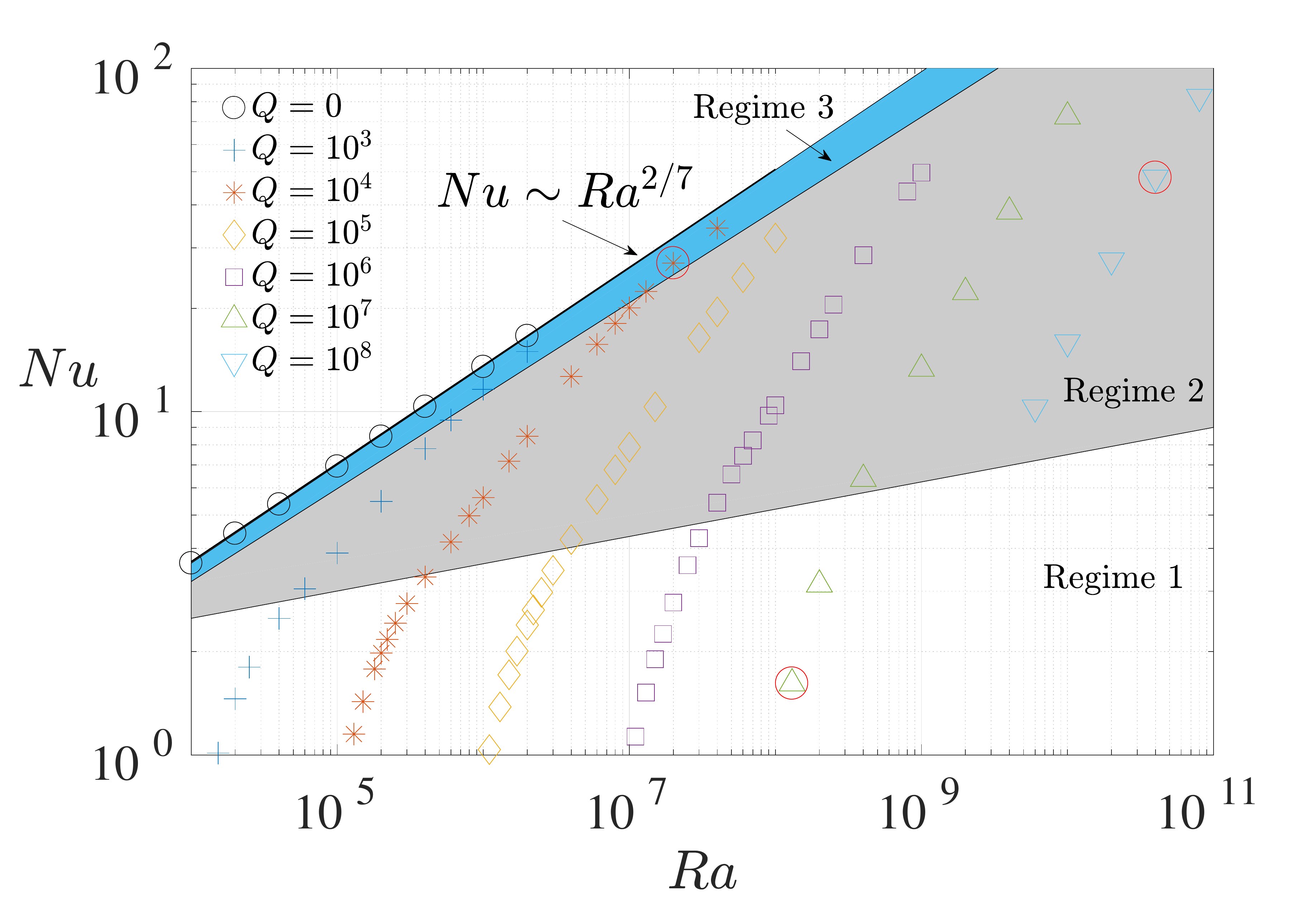}}  
    \subfloat[]{
      \includegraphics[height=5cm]{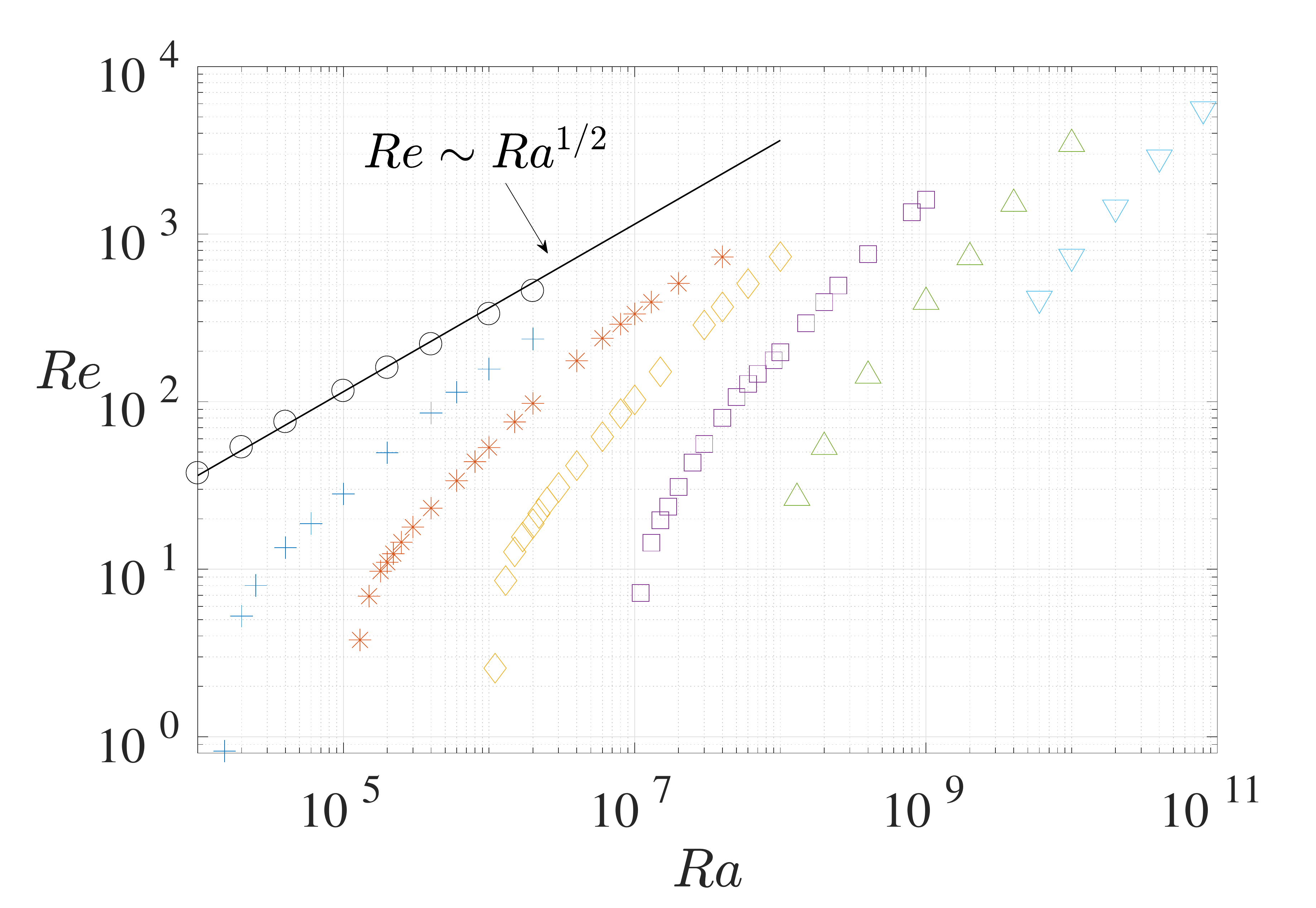}} 
            \qquad
    \subfloat[]{
      \includegraphics[height=5cm]{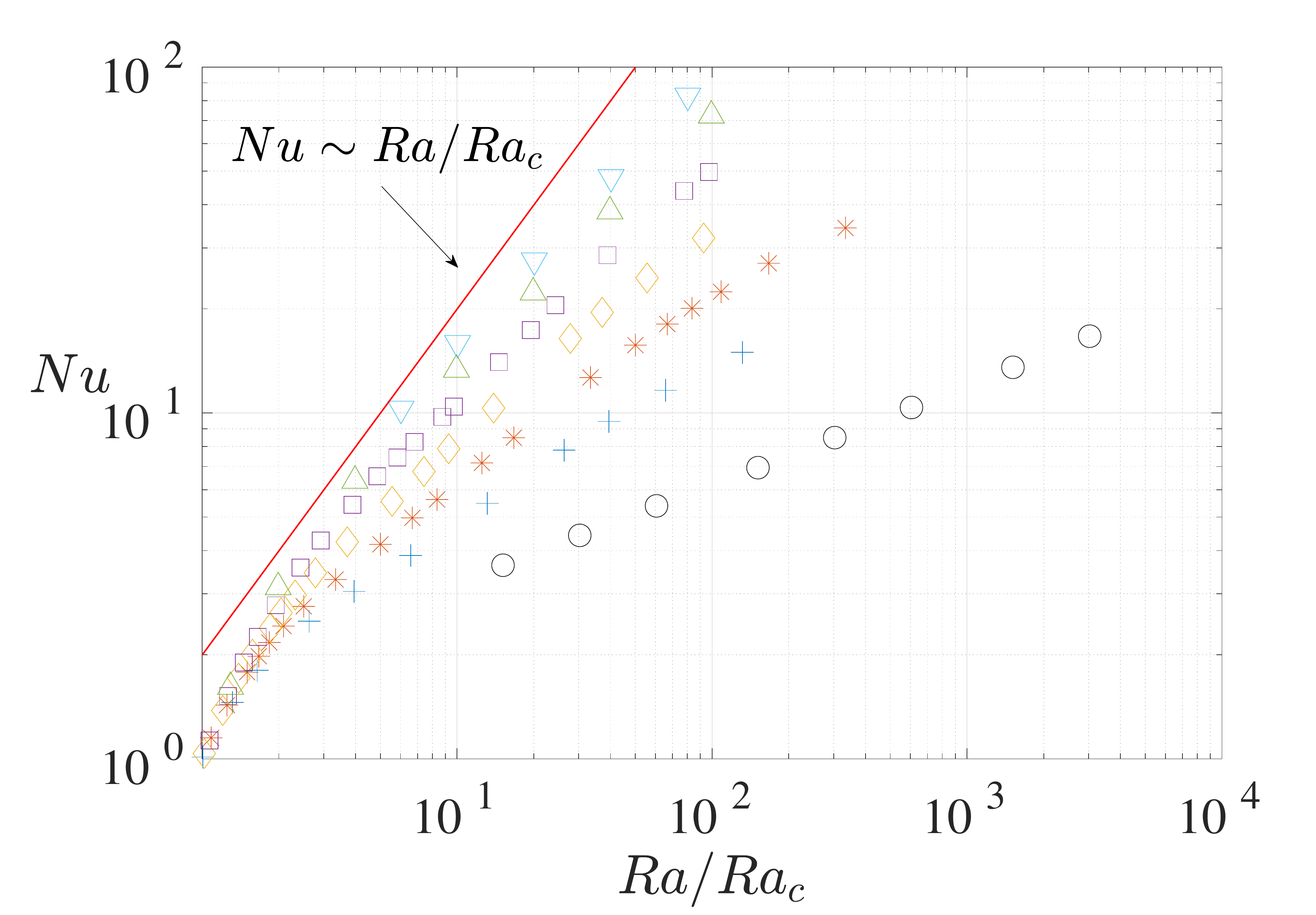}} 
          \subfloat[]{
      \includegraphics[height=5cm]{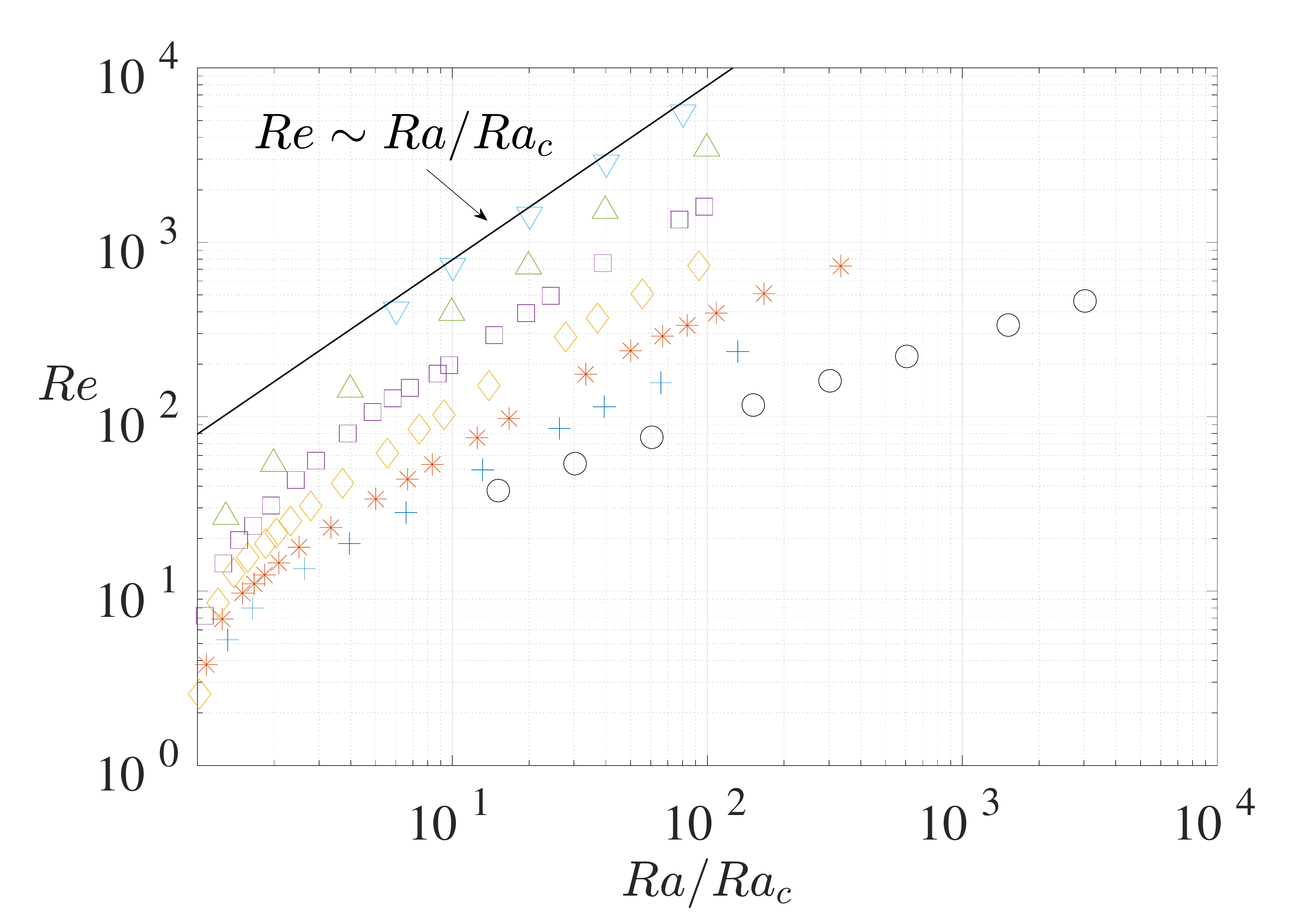}} 
  \end{center}
\caption{ Heat transport and flow speeds in magnetoconvection.  
(a) Nusselt number versus Rayleigh number (cases shown in Fig.~\ref{F:flow_regimes} and  Fig.~\ref{F:force} are marked with a red circle); (b) Reynolds number versus Rayleigh number; (c) Nusselt number versus $Ra/Ra_c$; (d) Reynolds number versus $Ra/Ra_c$. The three different regimes identified in the present study are illustrated in (a); regime 1 is the cellular regime, regime 2 is the columnar regime and regime 3 is the turbulent regime.}
\label{F:NuRa}
\end{figure*}

Figs.~\ref{F:NuRa}(a) and (b) show the Nusselt number ($Nu$) and Reynolds number ($Re$) versus the Rayleigh number ($Ra$) for all $Pr=1$ cases. The non-magnetic ($Q=0$) case is shown for comparison, along with the $Nu \sim Ra^{2/7}$ scaling typically found in studies using moderate $Ra$ and $Pr=O(1)$ \citep[e.g.][]{bC89} 
 and the `free-fall' scaling, $Re \sim Ra^{1/2}$. 
  All of the MC cases show qualitatively similar behavior to each other in their functional dependence of $Nu$ and $Re$ on $Ra$.  Figs.~\ref{F:NuRa}(c) and (d) show $Nu$ and $Re$ versus $Ra/Ra_c$, where the similarities between cases with different $Q$ values and  the asymptotic behaviors can be more clearly seen.

\begin{figure}
  \begin{center}
\subfloat[]{
      \includegraphics[height=4.5cm]{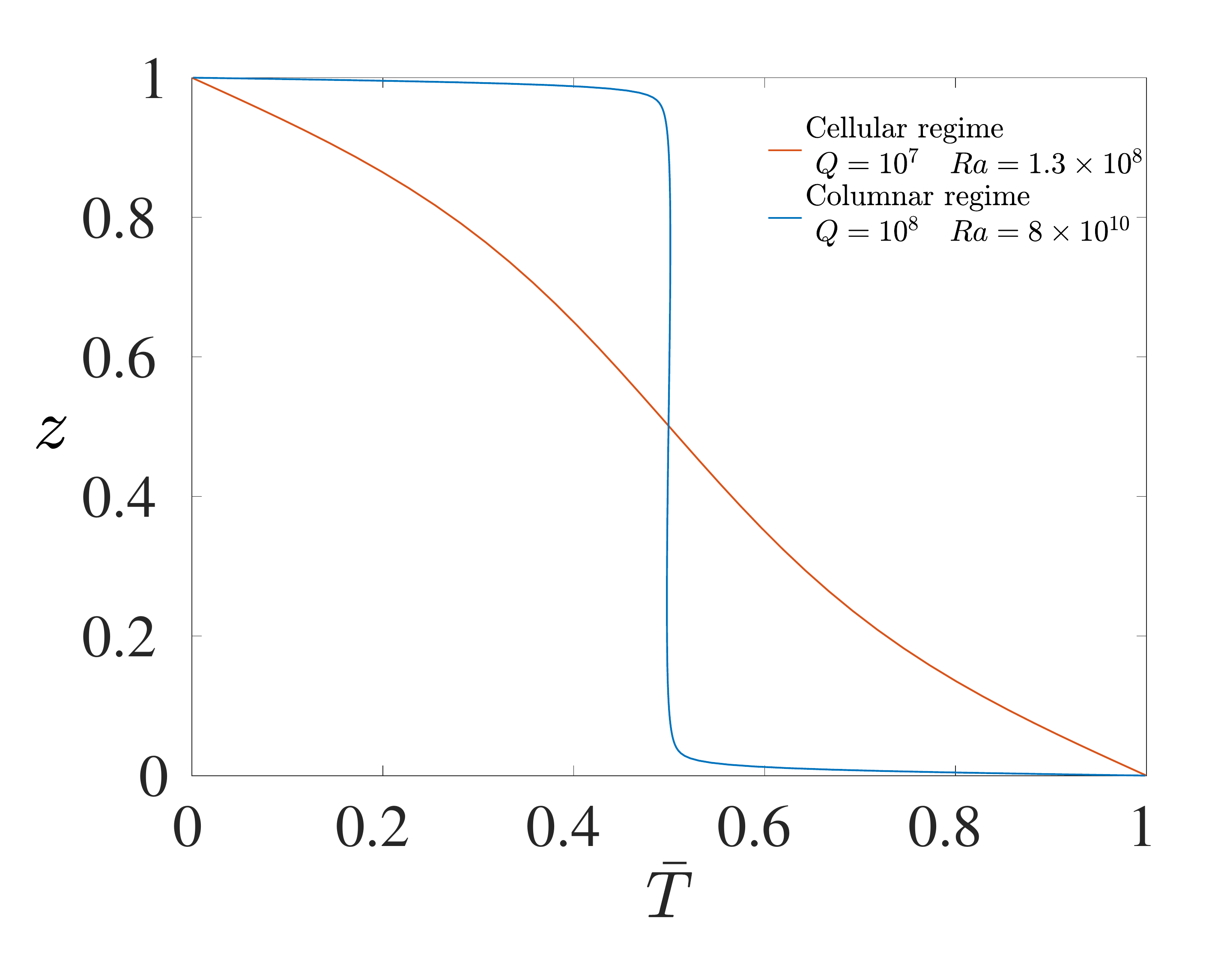}  }
      \subfloat[]{
      \includegraphics[height=4.5cm]{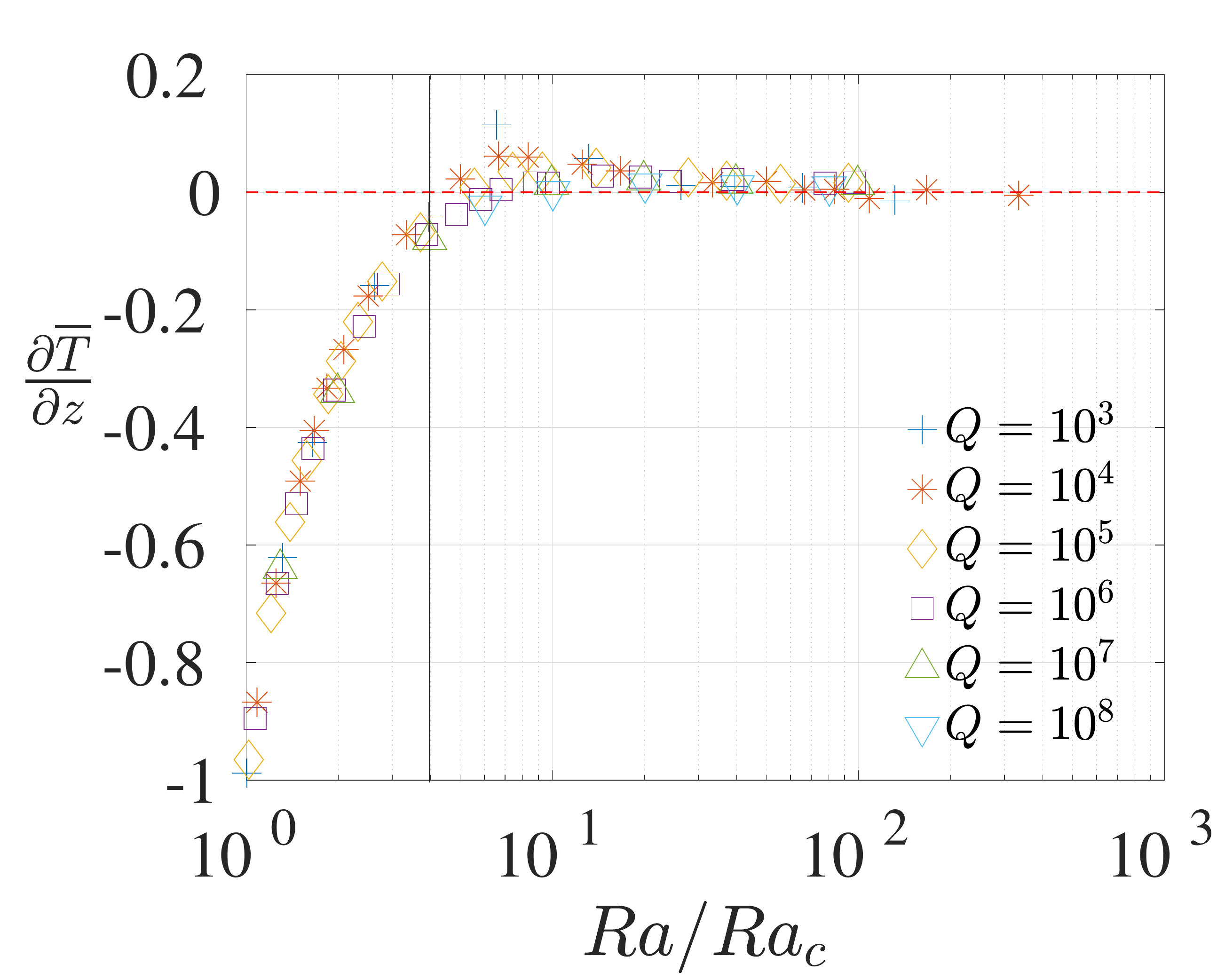}  }
  \end{center}
\caption{Characteristics of the time- and horizontally-averaged (mean) temperature ($Pr=1$). (a)  Mean temperature profiles for two representative cases within the cellular and columnar regimes. (b) Vertical gradient of the mean temperature at the vertical midplane. The solid black line roughly suggests the transition from the cellular regime to the columnar regime. The dashed red line indicates an isothermal interior.}
\label{F:tgrad}

\end{figure}

\begin{figure}
  \begin{center}
      \includegraphics[height=6cm]{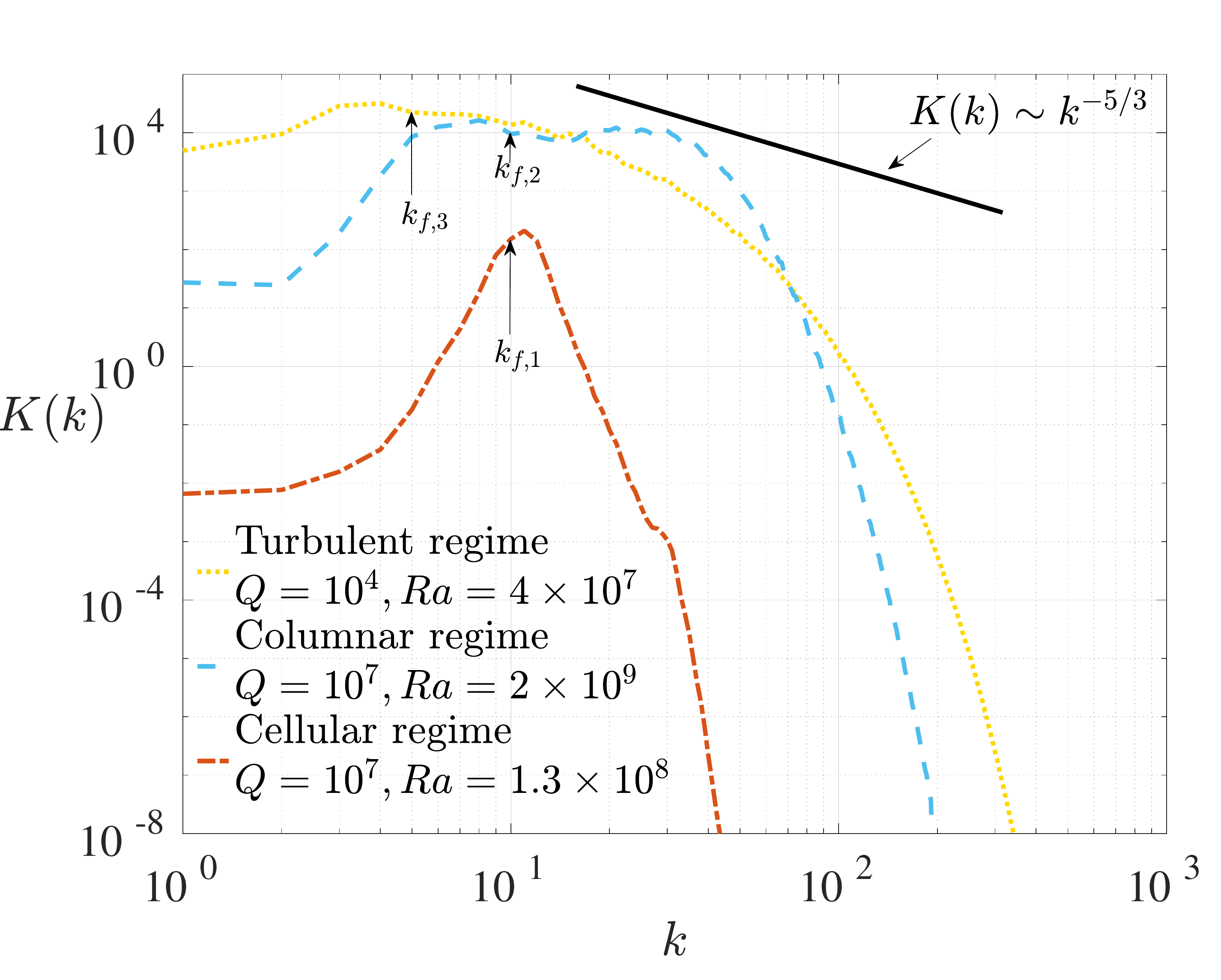}    
  \end{center}
\caption{Instantaneous kinetic energy spectra, $K(k)$, from representative cases of the three different magnetoconvection regimes; spectra are computed at the vertical midplane ($z=0$). The cellular regime is shown with the dash-dot line ($Q=10^7, Ra=1.3\times 10^8$); the columnar regime is shown with the dashed line ($Q=10^7, Ra=2\times 10^9$); and the turbulent regime is shown with the dotted line ($Q=10^4, Ra=4\times 10^7$).  
$k_{f,1}$,  $k_{f,2}$ and $k_{f,3}$ are the critical (forcing) wavenumbers for each case. The $k^{-5/3}$ slope is shown for reference.}
\label{F:spe}
\end{figure}

The first, cellular regime is characterized by the cellular structures reminiscent of linear convection, as illustrated in the visualizations of Fig.~\ref{F:flow_regimes}(a,b). In this regime, the heat transfer and flow speeds increase rapidly with increasing $Ra$, but with a s
slope that continuously decreases.
The time-averaged mean temperature profile for a typical case in this regime is shown in Fig.~\ref{F:tgrad}(a). Here, convective nonlinearities remain  weak and the characteristic scale of fluid motion remains dominated by the critical horizontal wavenumber, as illustrated in 
the kinetic energy spectra shown in Fig.~\ref{F:spe}.


When $Ra$ is increased to $\simeq4Ra_c$, the slope of the ($Nu$,$Ra$) and ($Re$,$Ra$) curves for each value of $Q$ appear to approach a constant value over a wide range of $Ra$. We refer to this regime as `columnar' because of the characteristic structure of the flow field shown in Fig. \ref{F:flow_regimes}(c,d), which  consists of a network of spatially-localized columns that span the depth of the layer. Flow speeds, as indicated by the Reynolds number in Fig.~\ref{F:NuRa}(b), become large in this regime in the sense that an equivalent $Q=0$ case would yield a turbulent flow, yet the fluid remains quasi-laminar and coherent. The simulations show that the columnar regime occupies an increasing range of $Ra$ as $Q$ is increased. The spatial localization of the convection leads to a locally-flattened kinetic energy spectrum centered near the critical wavenumber ($k = 10$ for the case shown), as illustrated in Fig.~\ref{F:spe}.   
Within the columnar regime, Figs.~\ref{F:tgrad}(a) and \ref{F:tgrad}(b) shows that the fluid interior becomes nearly isothermal and the thermal boundary layers are well-established.

\begin{figure*}
\begin{center}
\hspace*{-0.6cm}

  \includegraphics[width=0.98\textwidth]{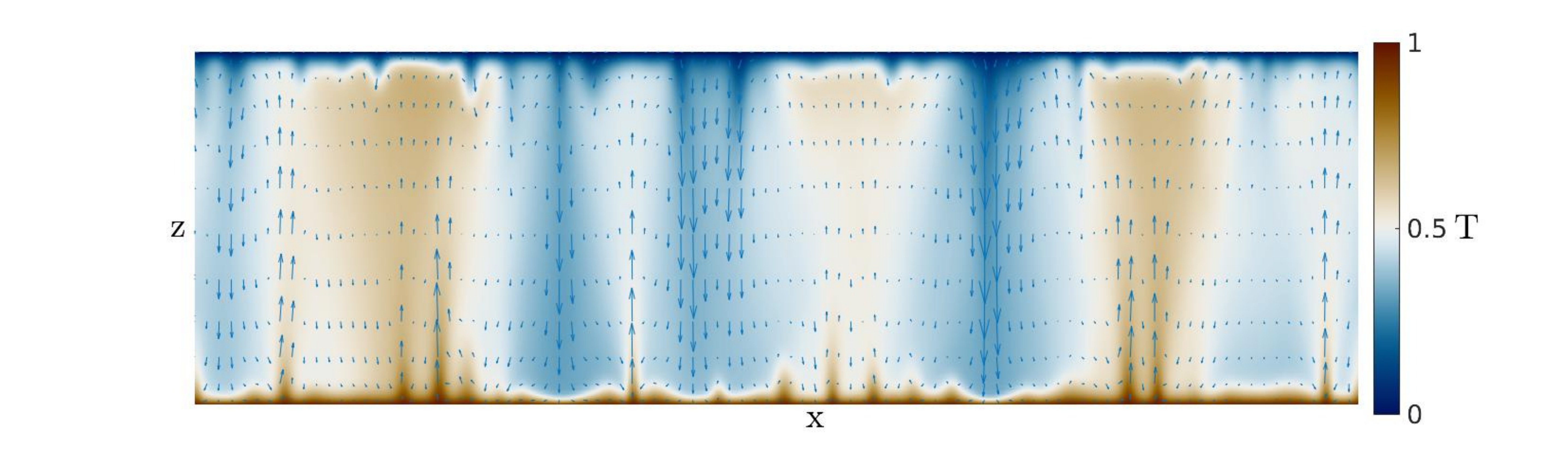}

 \end{center}
\caption{Two-dimensional ($x,z$) slice of temperature for a typical case in the magnetically-constrained columnar regime ($Q=10^6, Ra=1.5\times 10^8, Pr=1$), showing the broadening of thermal structures as hot (cold) fluid ascends (descends). The arrows indicate the velocity in the plane. }
\label{F:2DT}
\end{figure*}

As first observed by \cite{sC00}, the interior mean temperature gradient can be positive for some of the cases. A possible explanation for this reversed (stable) gradient is due to the vertical structure of the anisotropic columns, as illustrated in the vertical slice of the temperature in Fig.~\ref{F:2DT}. A given convection column exhibits an asymmetric structure about $z=0.5$; for instance, upwelling flow tends to be thin near the bottom boundary where flow converges, and broader near the upper boundary where flow diverges. Horizontally-averaging this flow structure results in a weakly-stable interior temperature profile. The columns act as an efficient heat transfer mechanism in the sense that heat is carried directly from boundary to boundary with limited horizontal mixing.

The third, turbulent MC regime is marked by a decrease in the slope of both the $Nu$ \citep[cf.][]{sC00} and $Re$ scalings with $Ra$. Here the columns disappear and, as a result, the flattened kinetic energy spectra observed in the columnar regime transition to a broader spectrum (Fig.~\ref{F:spe}) that is suggestive of a direct energy cascade and the development of an inertial subrange (the Kolmogorov spectrum $K(k) \sim k^{-5/3}$ is plotted for reference). However, the development of an inertial subrange is very slow and requires very large $Ra$. The Lorentz force still plays an important role in the dynamics, and extremely large Rayleigh numbers are required to leave the third regime and access flows in which the Lorentz force plays a negligible role.

\begin{figure*}
 \begin{center}
    \subfloat[]{
      \includegraphics[height=5cm]{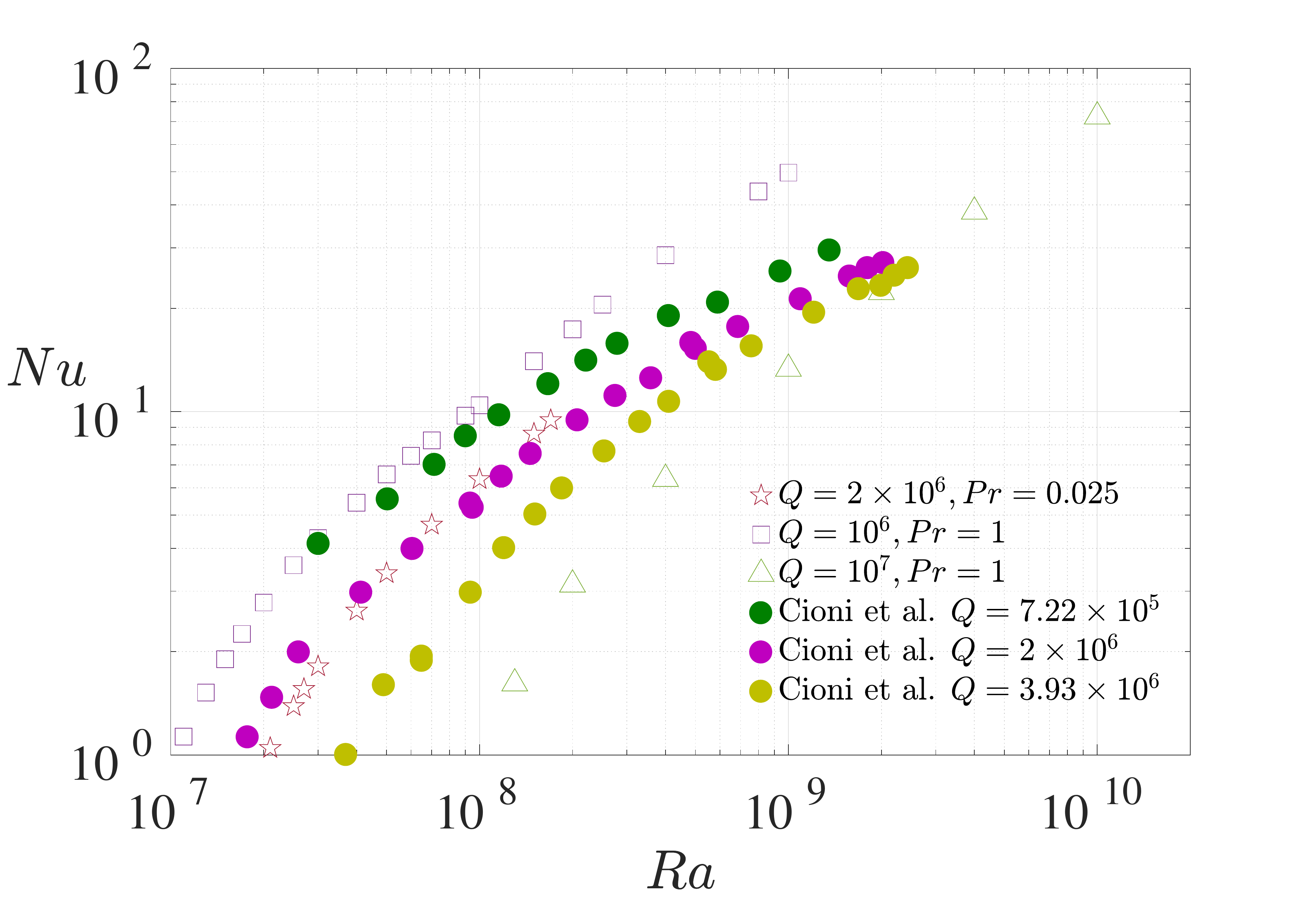}} 
          \subfloat[]{
      \includegraphics[height=5cm]{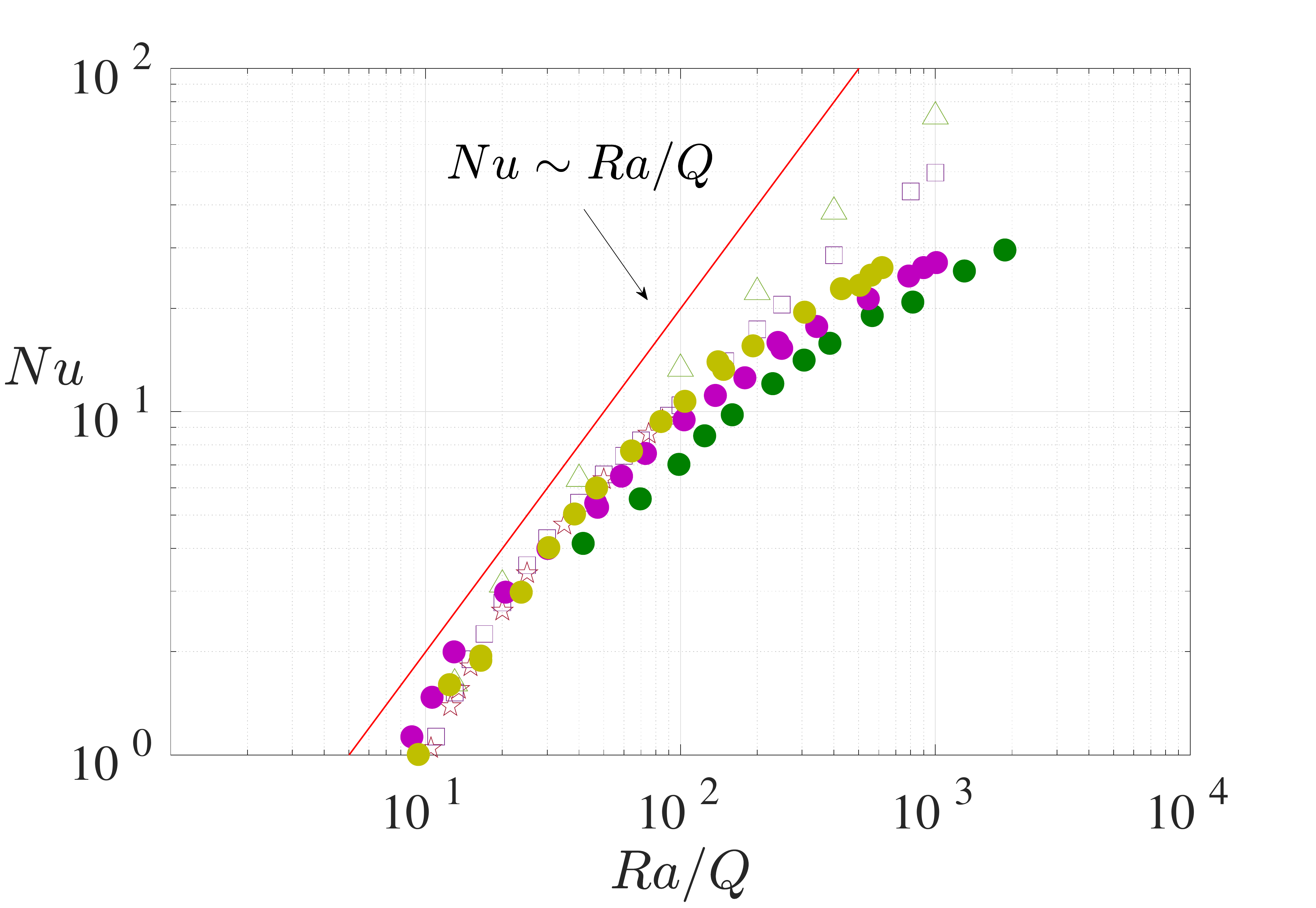}} 
  \end{center} 
\caption{Comparison of simulation results with the $Pr=0.025$ experimental data of \cite{sC00}. (a) Nusselt number versus Rayleigh number; (b) Nusselt number versus  $Ra/Q$. }
\label{F:Cioni}
\end{figure*}

\subsection{The influence of the Prandtl number}

A total of ten simulations with $Pr=0.025$ have also been carried out to determine the dynamical influence of the thermal Prandtl number, and to compare with available data from laboratory experiments that use this same value of $Pr$. Specifically, we use $Q=2 \times 10^{6}$ to compare with the results of \cite{sC00} who used a cylindrical container filled with mercury. Fig.~\ref{F:Cioni} shows the \cite{sC00} data along with the $Pr=0.025$ and two $Pr=1$ cases with comparable $Q$ values ($10^6$ and $10^7$)  from the present work. In general, similar behavior is seen between the simulations and the experimental data, despite the differences in boundary conditions and geometry.  As revealed by our results, the regime I of \cite{sC00} actually consists of two distinct magnetically-constrained regimes (cellular and columnar) in which the flow cannot accurately be described as turbulent. Although they suggest a $Nu \sim Ra/Q$ fit to their data, our simulations show that this fit only arises at much higher values of $Q$. The most significant discrepancies between the two datasets appear at higher values of $Ra$ (or $Ra/Q$), where the \cite{sC00} data shows a lower slope in comparison with the simulations (both $Pr=0.025$ and $Pr=1$). Because this difference is largest at higher values of $Ra$, it might be due to, as suggested by \cite{sC00}, the formation of a large-scale circulation in the experimental apparatus due to the eventual loss of magnetic constraint and flow anisotropy \citep[][]{tV18b,zL19}. Moreover, at the highest values of $Ra$ accessed by \cite{sC00} it is likely that the dynamics are within the third regime, within which different values of $Pr$ might lead to different scaling behavior. Additional studies using the cylindrical geometry and higher values of $Ra$ are necessary to quantify this effect in more detail.

\begin{figure*}
 \begin{center}

          \subfloat[]{
            \includegraphics[height=5.5cm]{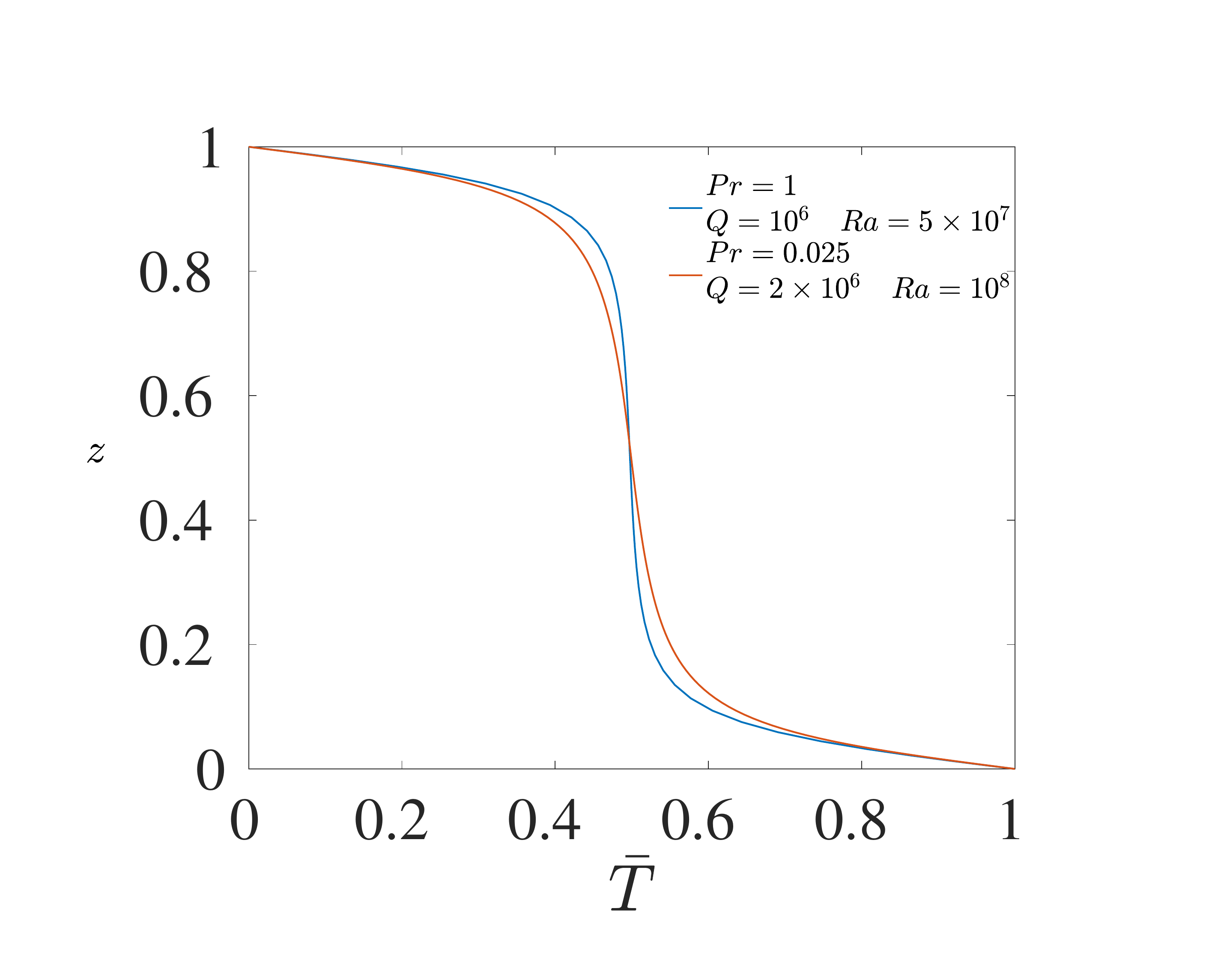}} 
                \subfloat[]{
      \includegraphics[height=5.5cm]{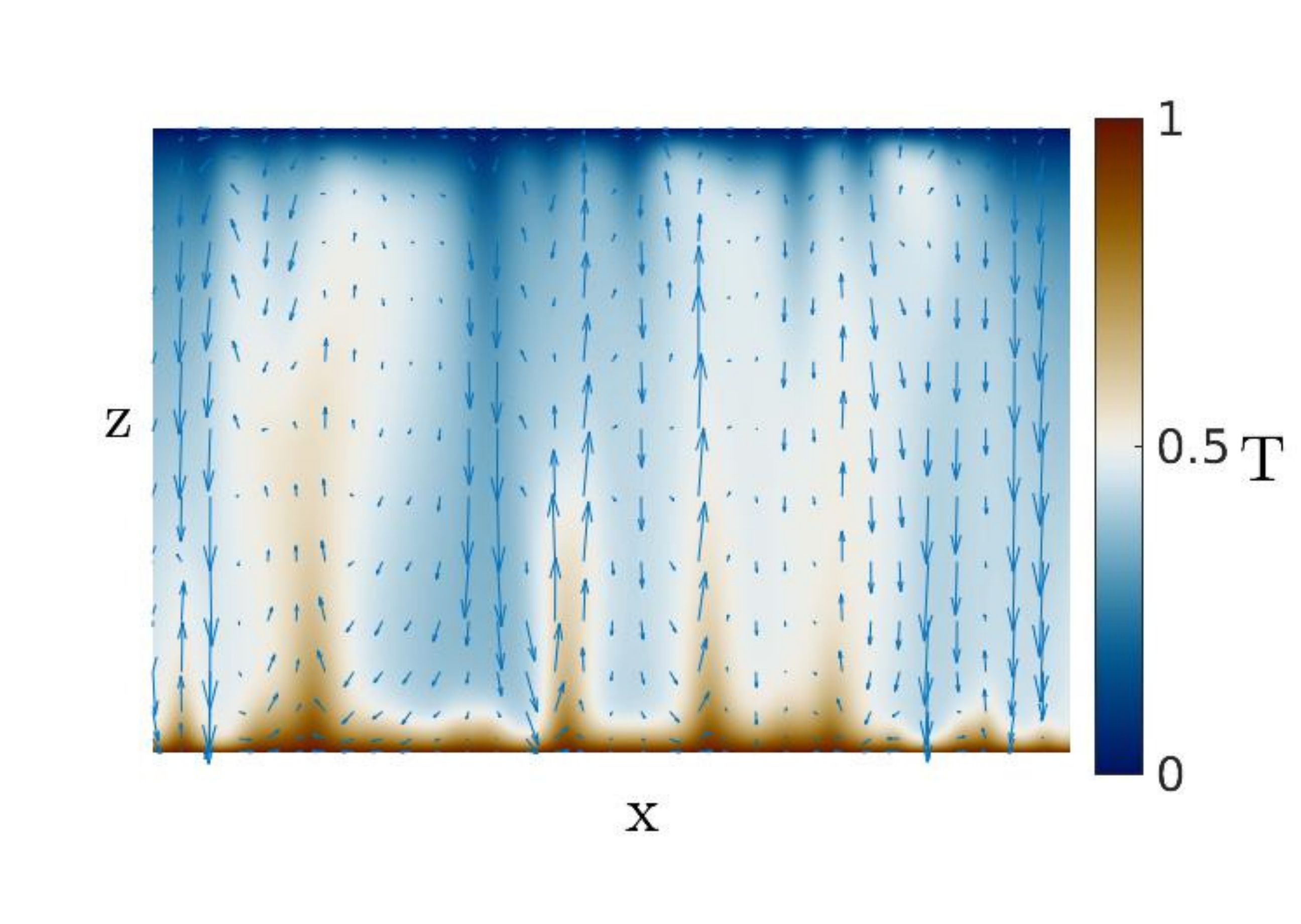}} 
  \end{center} 
\caption{The influence of the thermal Prandtl number on magnetoconvection.  (a) time- and horizontally-averaged (mean) temperature profile for cases with different $Pr$ but the same $Ra/Q$; (b) two-dimensional ($x,z$) slices of temperature  for columnar regime with $Pr=0.025, (Q=2\times 10^6, Ra=1.7\times 10^8)$. }
\label{F:NuRa1}
\end{figure*}

We find that the $Nu$-$Ra$ scaling behavior for $Pr=0.025$ is nearly identical to that observed for the $Pr=1$ cases, as shown in Fig.~\ref{F:Cioni}(b). In addition, we can identify the cellular and columnar regimes for the $Pr=0.025$ cases based on the $Nu$ scaling and flow structures. We note that single-mode MC dynamics are also $Pr$-independent \citep{pM99,kJ99c}. The mean temperature profiles in Fig.~\ref{F:NuRa1}(a) show that the thickness of the thermal boundary layers observed in the $Pr=0.025$ cases is comparable to the corresponding cases with $Pr=1$ for a given value of $Ra/Q$. However, a larger interior temperature gradient is present for the $Pr=0.025$ cases. Fig.~\ref{F:NuRa1} shows a two-dimensional ($x,z$) slice of the temperature field for a case in the columnar regime with $Pr=0.025$. Both $Pr=0.025$ and $Pr=1$ cases show similar anisotropic columnar structures. However, as might be expected with a smaller $Pr$, heat diffuses to the surrounding fluid  more rapidly  before it is carried to the top of the layer. As a result, cases with a smaller $Pr$  are expected to  require a larger $Ra/Ra_c$ to reach an isothermal interior for finite values of $Q$.

.eps

\subsection{Balances}

\begin{figure*}
\begin{center}
\includegraphics[height=6cm]{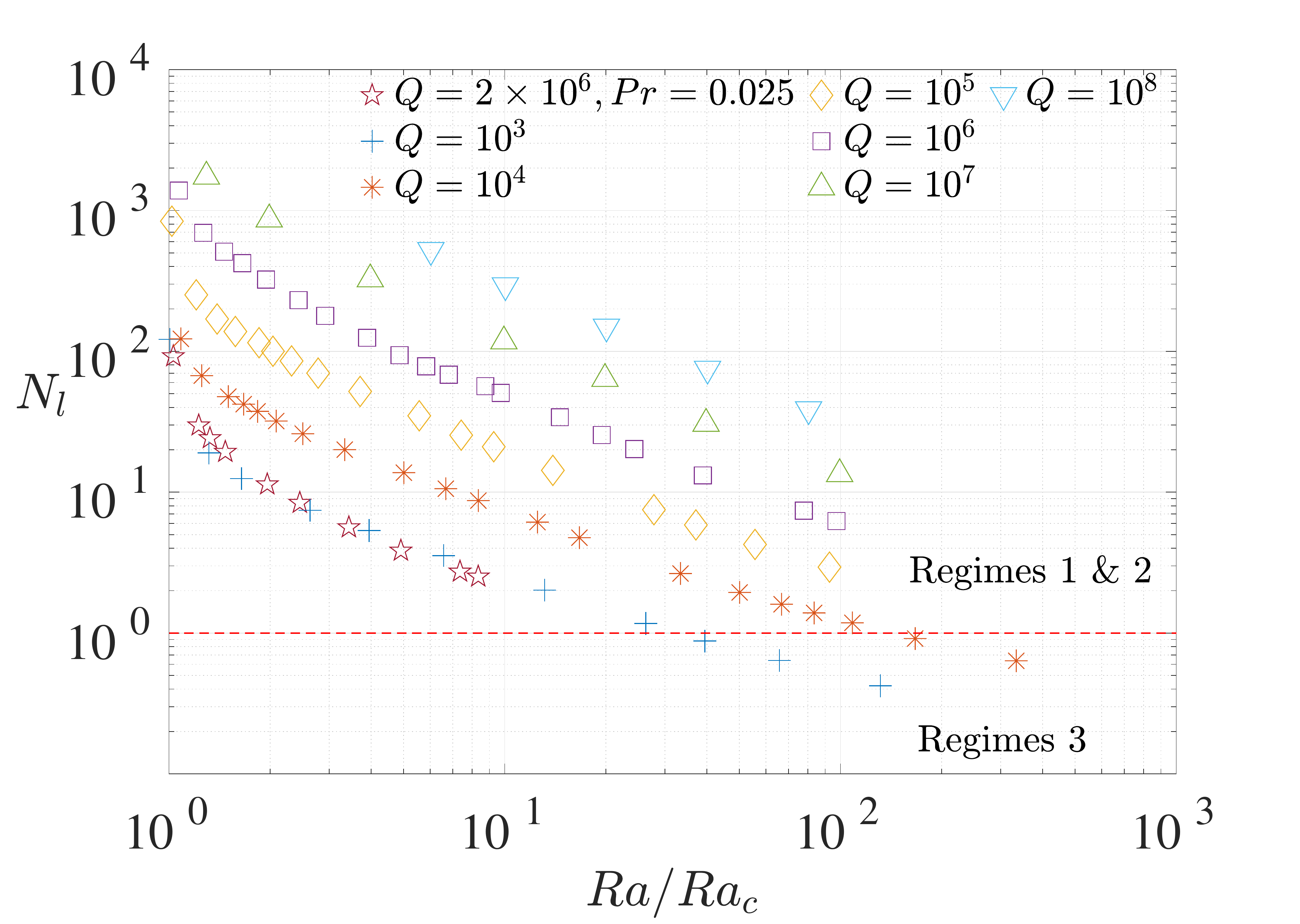} 
 \end{center}
\caption{Rescaled interaction parameter $N_\ell =Q^{2/3}/Re$ versus $Ra/Ra_c$. The Lorentz force remains dominant provided $N_\ell \gtrsim1$. The  dashed red line shows the approximate location for the transition from the columnar regime to the turbulent regime. Unless explicitly stated all curves are for $Pr=1$. }
\label{F:NRaQ_all}
\end{figure*}

The magnetic interaction parameter, $N = \mathcal{B}^2 L/(\rho \mu \eta U)$ (where $L$ and $U = |\ub|$ are generic characteristic length and speed scales) can be used to estimate the relative magnitudes of the Lorentz force to the advection terms in the momentum equation \citep[e.g.][]{pD13}. If one assumes that $L\equiv H$, we have $N = Q/Re$ \citep[e.g.][]{sC00}. In the present work, however,  a definition that better captures the transition between the regimes identified above can be found by incorporating the $Q$-dependent horizontal length scale of the convective flows into the definition of the interaction parameter; we denote this rescaled interaction parameter by $N_\ell$. From the vertical component of the momentum equation, for instance, we have 
\be
N_\ell \sim \frac{\text{Lorentz force}}{\text{advection}} \sim\frac{|Q \dsz b_z |}{|u_z \dsz u_z |} \sim Q \frac{| b_z | }{|u_z|^2} \sim Q \frac{|\ell^2 |}{|u_z|} \sim \frac{Q^{2/3}}{Re} ,
\ee
where, from the vertical component of the induction equation, we have used
\be
b_z = - \nabla^{-2} \dsz u_z \quad \Rightarrow \quad |b_z| \sim \ell^2 |u_z| = \ell^2 Re .
\ee
We assume $\ell \sim Q^{-1/6}$, vertical derivatives are of order one, and the vertical component of the velocity is asymptotically-larger than the horizontal components so that $u_z \sim Re$ \citep[e.g.][]{pM99}. An identical relationship can be found from the horizontal components of the momentum equation by relating the horizontal and vertical velocity components through the continuity equation. Thus, the transition from the columnar regime to the turbulent regime is expected when $N_\ell \simeq 1 $. Fig.~\ref{F:NRaQ_all} shows the interaction parameter $N_\ell$ plotted versus $Ra/Ra_c$. The dashed red line indicates the boundary between the columnar and turbulent regimes. It is suggested that with a larger $Q$ value, a larger $Ra/Ra_c$ is required to leave the columnar regime. The $Pr=0.025$ data suggests that smaller values of $Pr$ will yield a narrower (in terms of Ra) columnar regime, and a broader turbulent regime for a fixed value of $Q$. 

\begin{figure*}
  \begin{center}
    \subfloat[]{
     \includegraphics[height=3.5cm]{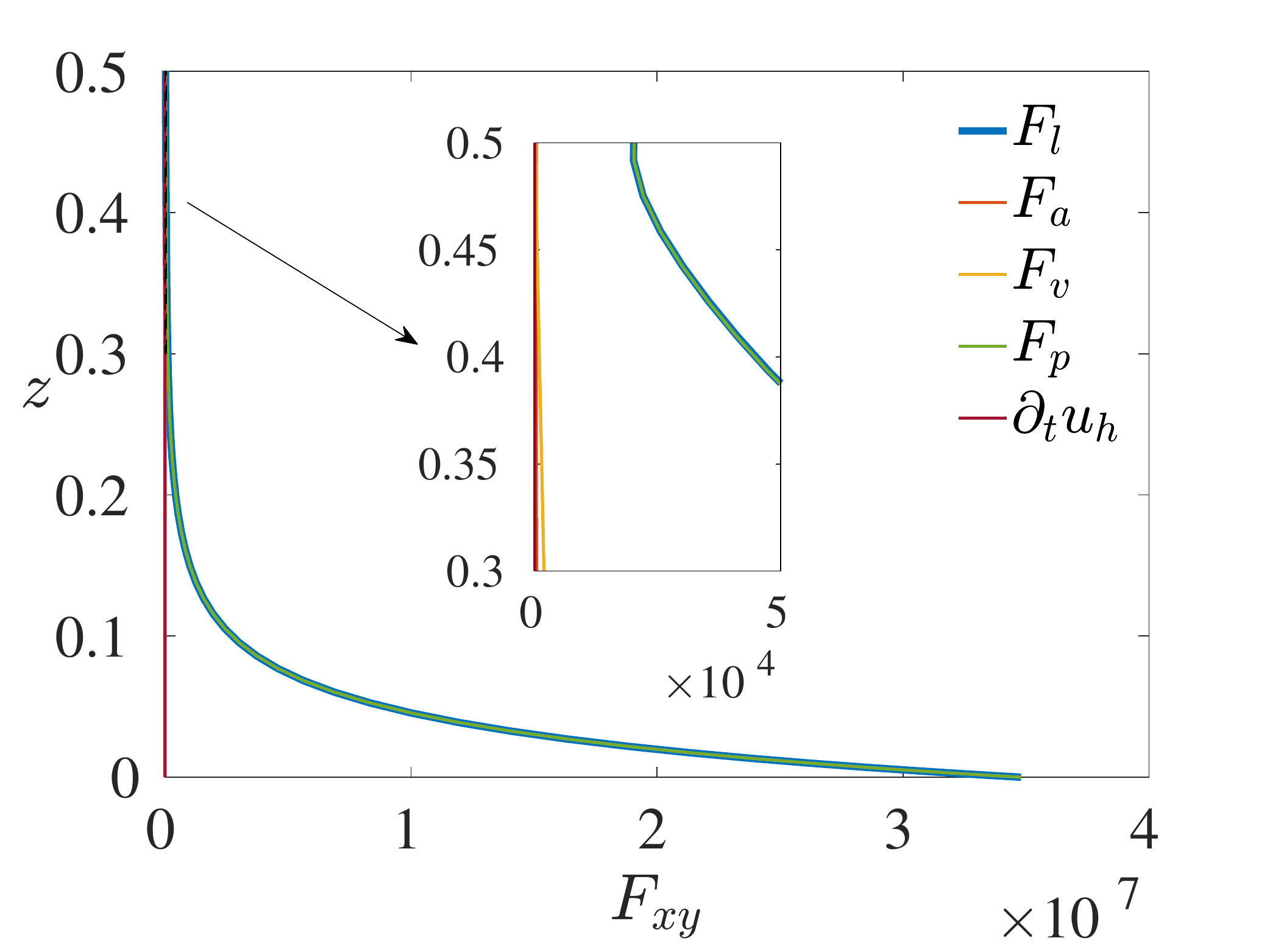}}
       \subfloat[]{ 
       \includegraphics[height=3.5cm]{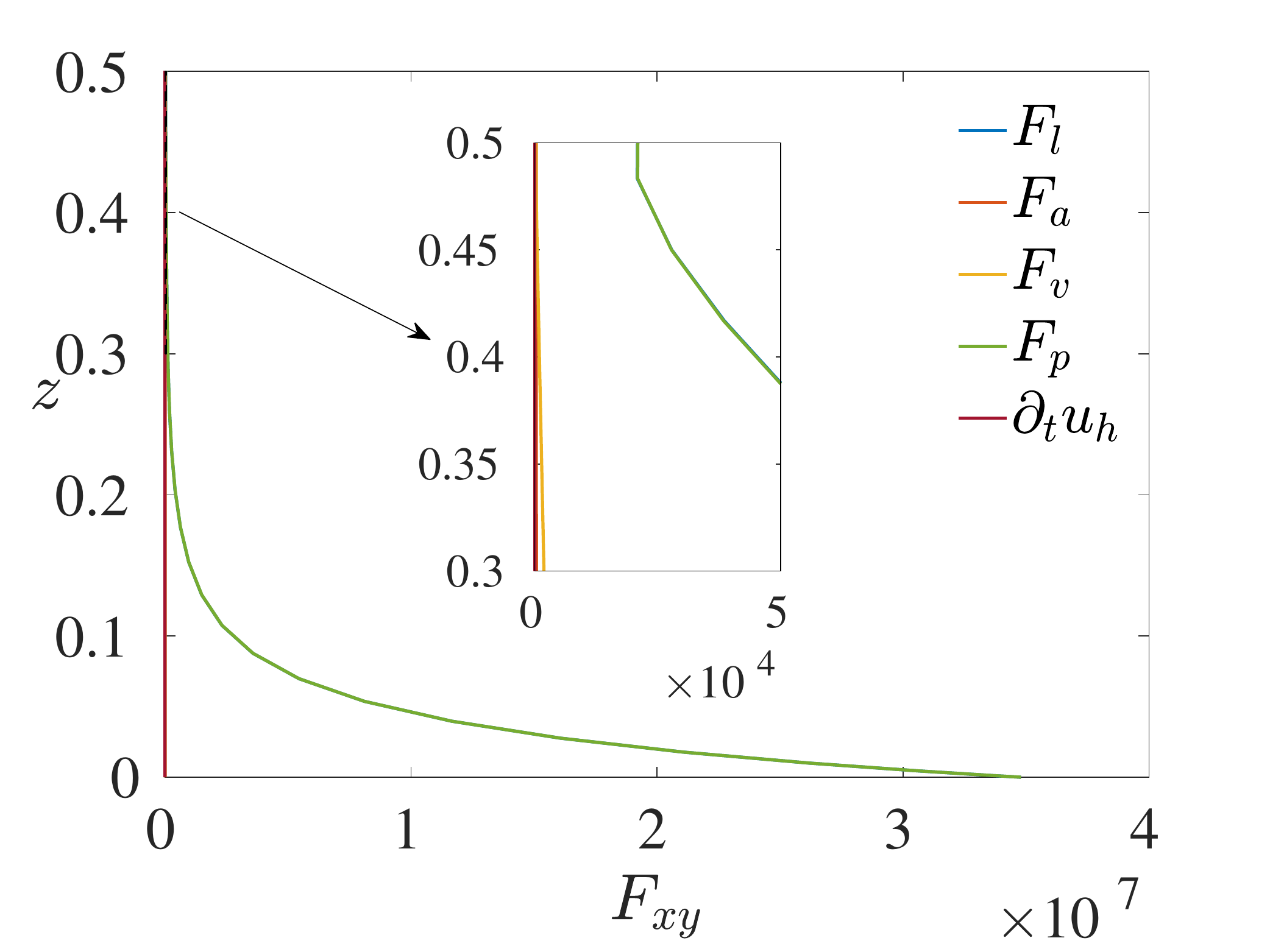}}
         \subfloat[]{ 
             \includegraphics[height=3.5cm]{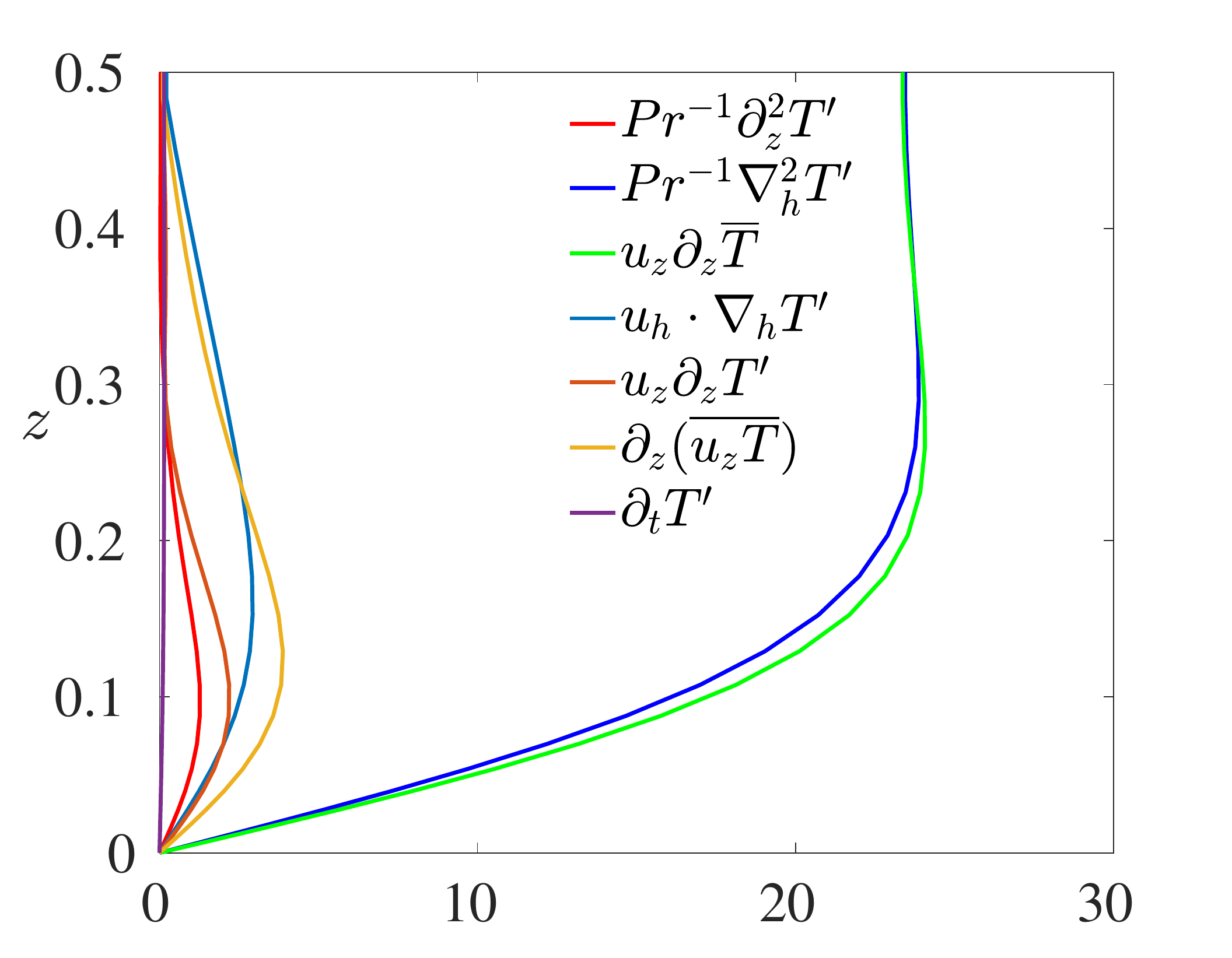}} 
       \qquad
  \subfloat[]{
     \includegraphics[height=3.5cm]{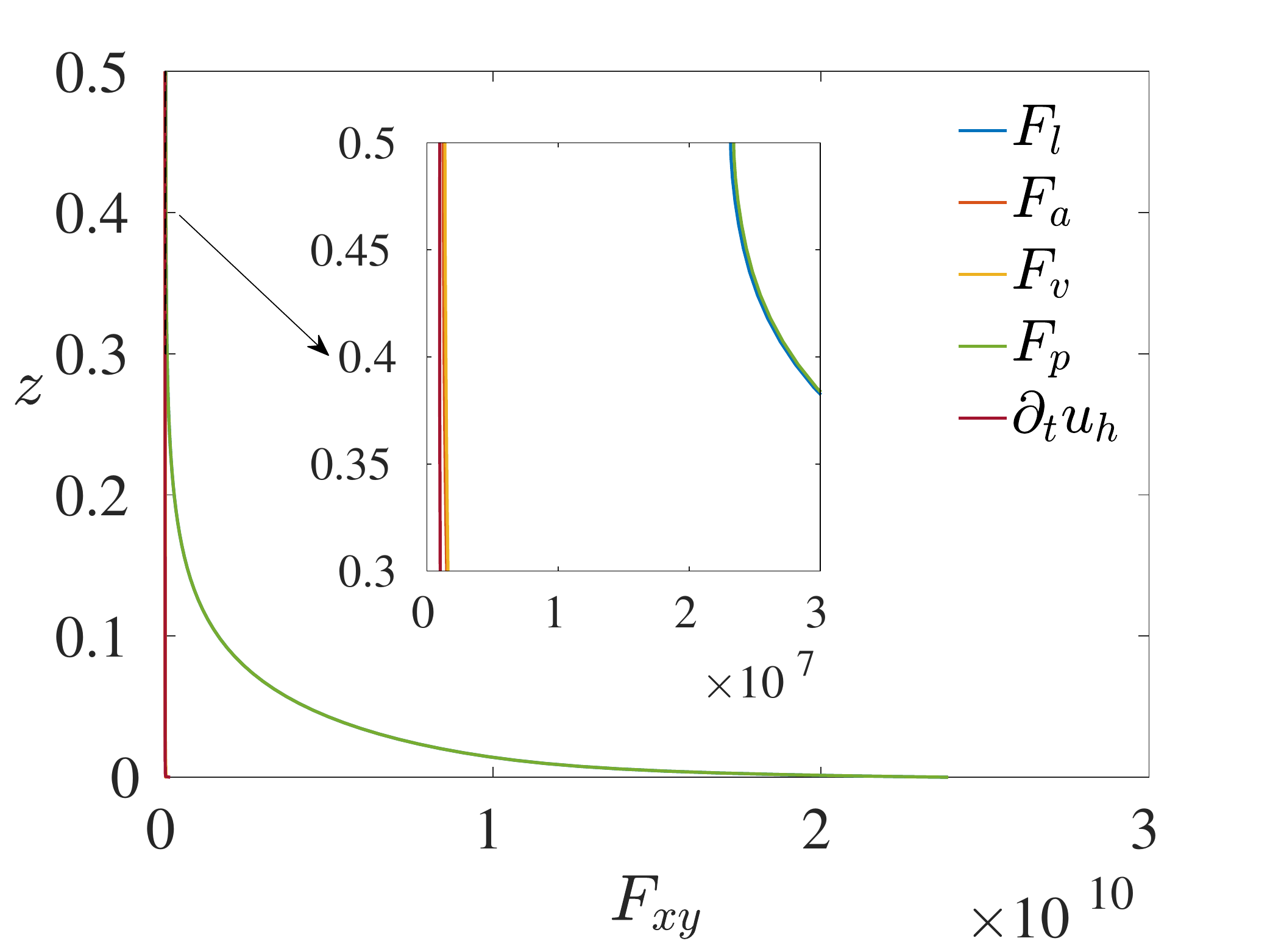}}
       \subfloat[]{ 
       \includegraphics[height=3.5cm]{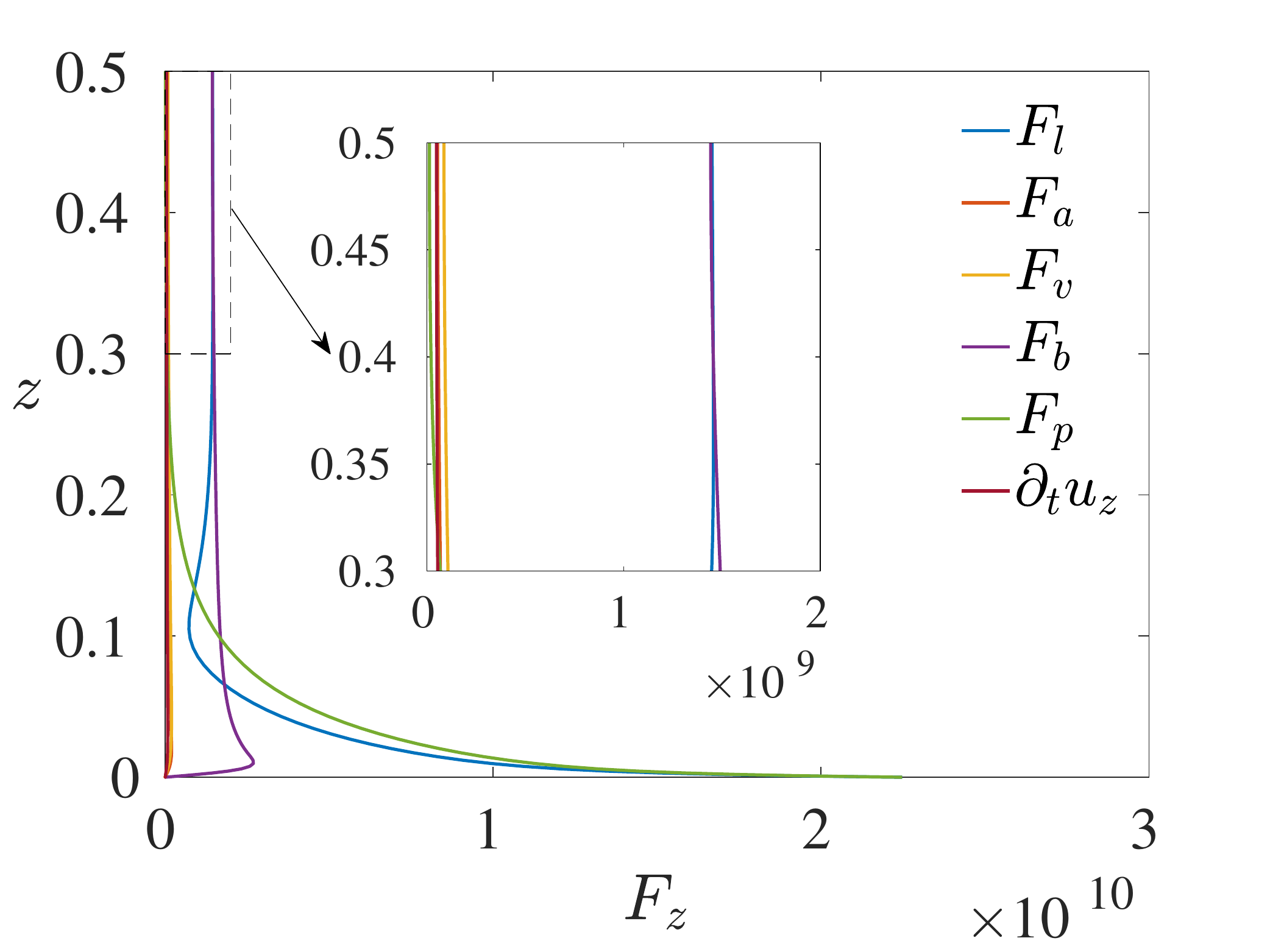}}
         \subfloat[]{ 
             \includegraphics[height=3.5cm]{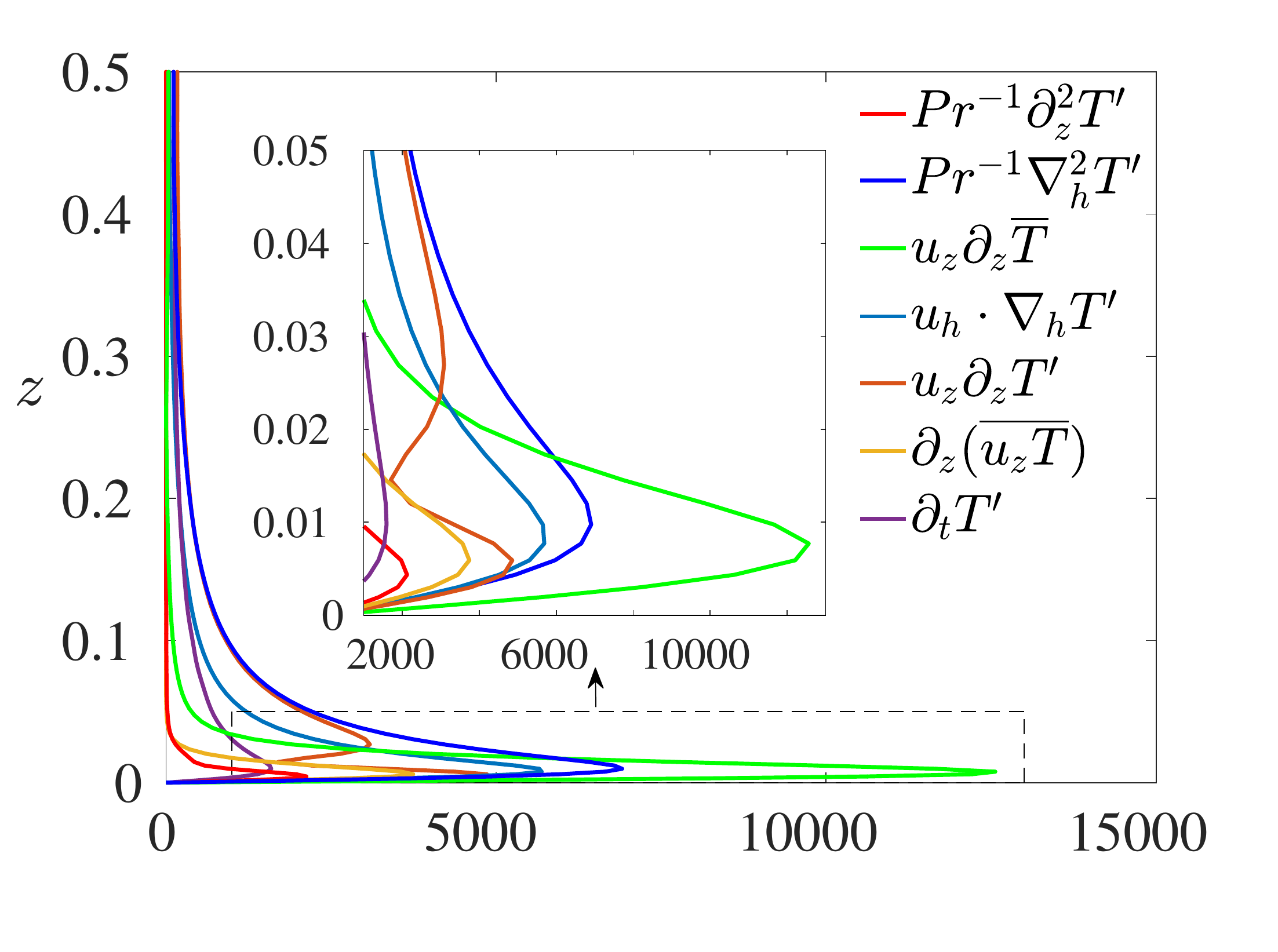}} 
       \qquad
    \subfloat[]{
     \includegraphics[height=3.5cm]{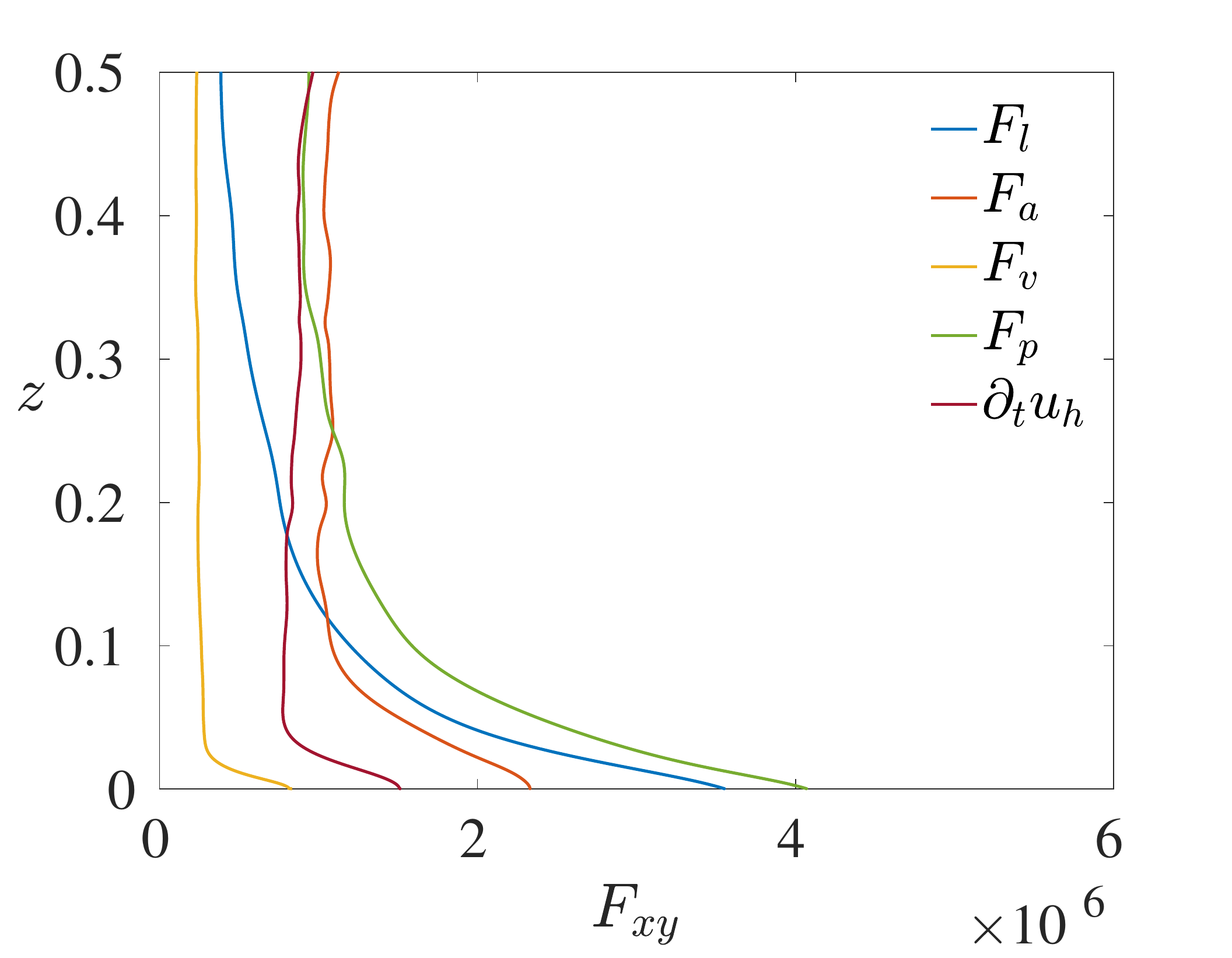}}
       \subfloat[]{ 
       \includegraphics[height=3.5cm]{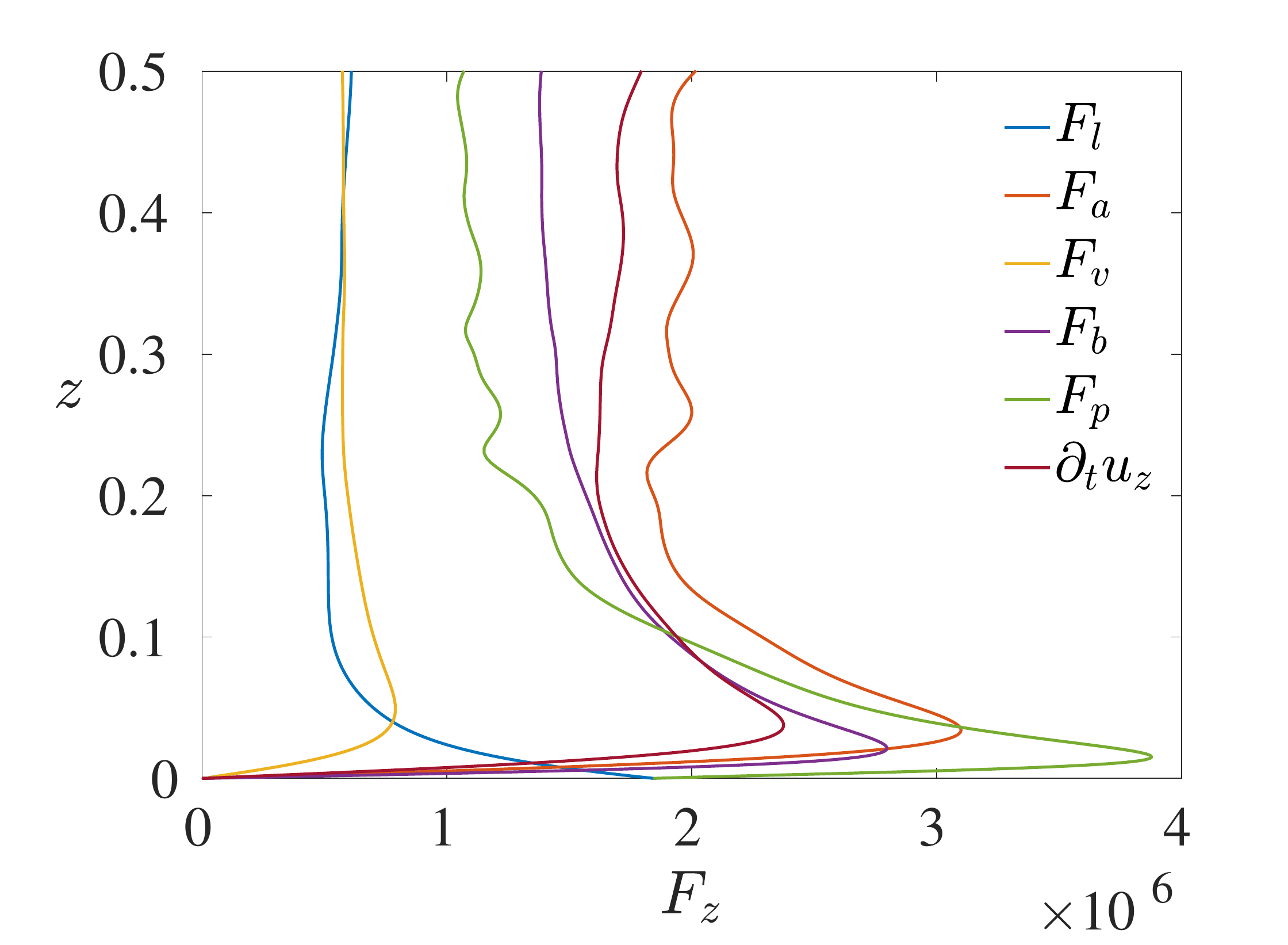}}
         \subfloat[]{ 
             \includegraphics[height=3.5cm]{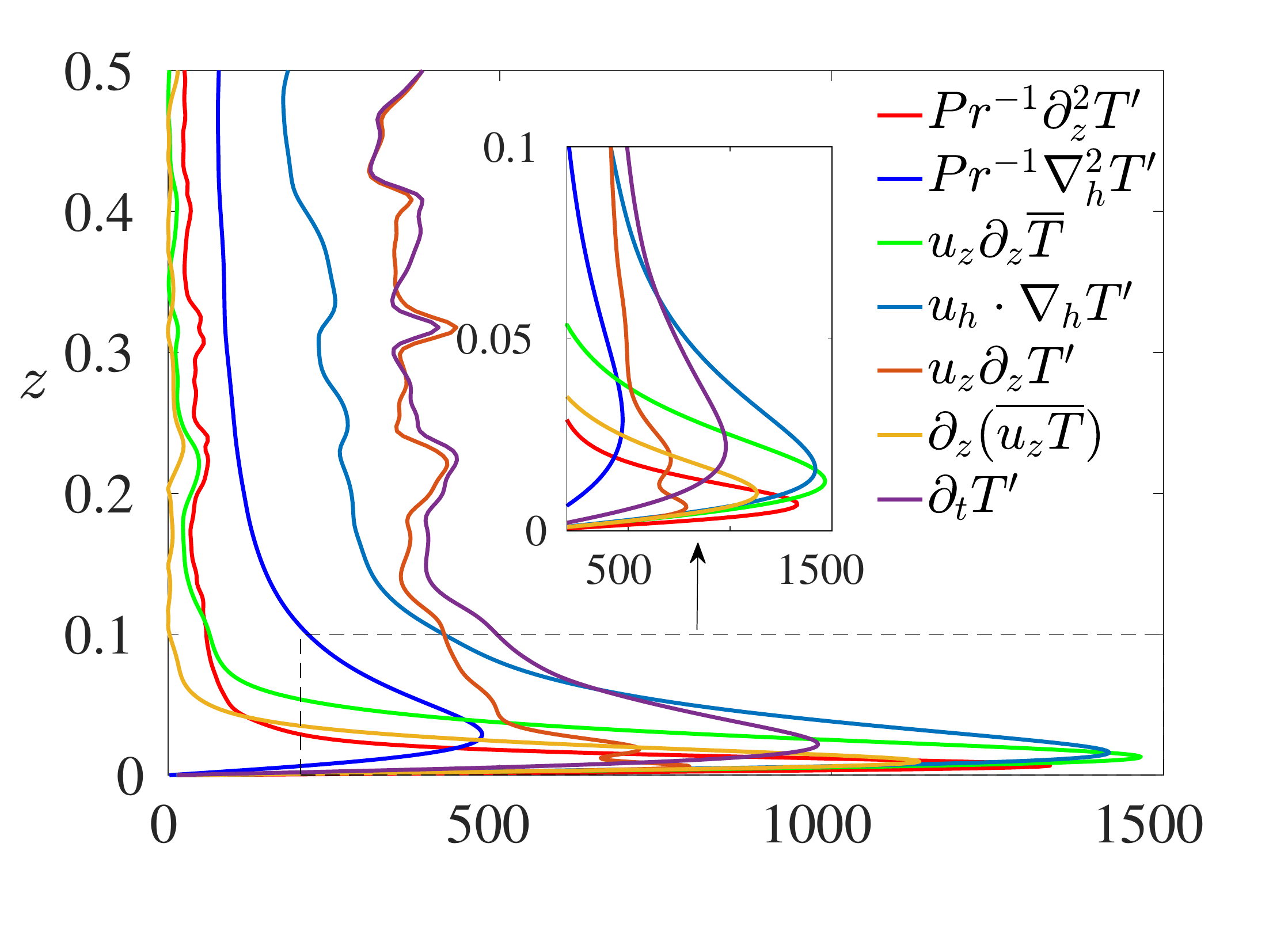}}  
                    \qquad      
  \end{center}
\caption{Instantaneous dynamical balances in $Pr=1$ magnetoconvection. Vertical profiles of the horizontal rms of each term present in the momentum  equation (\ref{eq7}) [(a),(b),(d),(e),(g),(h)] and fluctuating heat equation  (\ref{eq9})  [(c),(f),(i)] are shown.
The cellular regime ($Q=10^7$, $Ra=1.3\times10^{8}$): (a) horizontal forces $F_{xy}=(F_x^2+F_y^2)^{1/2}$; (b) vertical forces $F_z$; (c) all terms in the fluctuating heat equation.   The columnar regime ($Q=10^8$, $Ra=4\times10^{10}$): (d) horizontal forces $F_{xy}$; (e) vertical forces $F_z$; (f) all terms in the fluctuating heat equation. The turbulent regime ($Q=10^4$, $Ra=2\times10^7$): (g) horizontal forces; (h) vertical forces; (i) all terms in the fluctuating heat equation. Vertical velocity, horizontal velocity, and horizontal gradient are given by $u_z$, $u_h=(u_x^2+u_y^2)^{1/2}$ and $\nabla_h=(\partial_x,\partial_y,0)$.  Advection and the Lorentz, viscous, buoyancy and pressure gradient forces are denoted by $F_a$, $F_l$, $F_v$, $F_b$ and $F_p$, respectively. Inertia in the horizontal and vertical components of the momentum equation are denoted by $\partial_t u_h$ and $\partial_t u_z$, respectively.}
\label{F:force}
\end{figure*}

Fig. \ref{F:force} shows vertical profiles of the instantaneous horizontal rms of each term in the governing equations, for representative $Pr=1$ cases in the cellular [(a)-(c)],  columnar [(d)-(f)] and turbulent [(g)-(i)] regimes. Instantaneous values of the profiles are used to demonstrate that the force balances in the magnetically-constrained regimes apply at all times, and show remarkably smooth profiles; therefore, the choice of the particular instant in time does not affect the results. The interior and boundary layer are characterized by distinct balances, so it is helpful to consider the two regions separately. 
As predicted from the linear asymptotic scalings  \citep[e.g.][]{sC61,pM99}, Figs. \ref{F:force}(a), (b), (d) and (e) show that, for the cellular and columnar regimes, the leading-order force balance within the interior is between the Lorentz force ($F_l$), buoyancy force ($F_b$) and horizontal pressure gradient ($F_p$), 

\begin{linenomath*}
\be
0 \approx Q \partial_z b_z + \frac{Ra}{Pr}T^\prime, \quad0 \approx Q \partial_z b_h  - \nabla_\perp\Pi \quad \textnormal{(Regimes\; 1\; and \;2,\; interior)} .
\ee
\end{linenomath*}
  In the boundary layers we find that the pressure gradient balances the Lorentz force, 
  \begin{linenomath*}
\be
0 \approx Q \partial_z \bold{b}   - \nabla\Pi,\quad 
\textnormal{(Regimes\; 1\; and \;2,\; boundary layers)}
\ee
\end{linenomath*}
For the cellular regime, the vertical advection of the mean temperature and the horizontal diffusion terms are in dominant balance in the entire flow domain \citep[cf.][]{pM99},
  \begin{linenomath*}
\be
 u_z\partial_z\overline{T} \approx Pr^{-1} \nabla_\perp^2 T^\prime . \quad \textnormal{(Regime\; 1)} 
\ee
\end{linenomath*}

For the columnar regime, in the interior, the dominant terms in the fluctuating heat equation  (\ref{eq9}) are 
\begin{linenomath*}
\be
\dst T' + u_z \dsz T' \approx Pr^{-1} \nabla_\perp^2 T' ,\quad \textnormal{(Regime\; 2, \;interior)}
\ee
\end{linenomath*}
whereas vertical advection of the mean temperature dominates in the boundary layers. 
The balance in the interior suggests that for a fixed value of $Q$, the horizontal length scale of the columns is, in addition to being strongly dependent upon $Q$, also limited by horizontal thermal diffusion.

In the turbulent regime [Fig.~\ref{F:force}(d), (e) and (f)], the simulations show that advection, inertia, and the pressure gradient and buoyancy forces are important throughout the fluid layer.
 The turbulent regime possesses no instantaneous force balance in the sense that inertia and advection are dominant, whereas the Lorentz and viscous forces are subdominant. The heat equation is dominated by nonlinear advection within the interior, and all terms but horizontal diffusion play a significant role in the thermal boundary layer.

\begin{figure*}
\begin{center}
         \subfloat[]{ 
         \hspace*{-0.6cm} 
             \includegraphics[height=5.5cm]{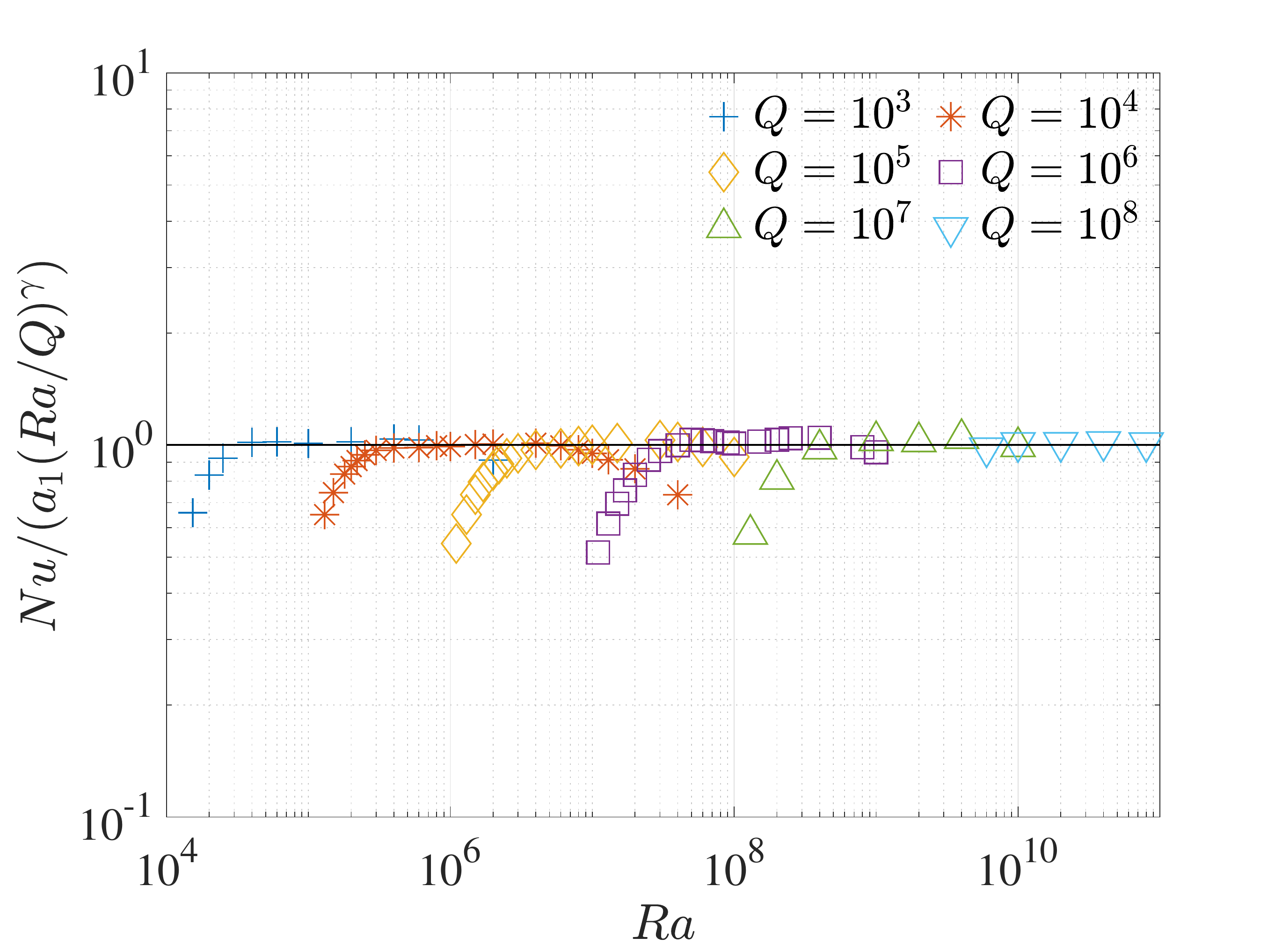}} 
      \hspace*{-0.8cm} 
        \subfloat[]{ 
\includegraphics[height=5.5cm]{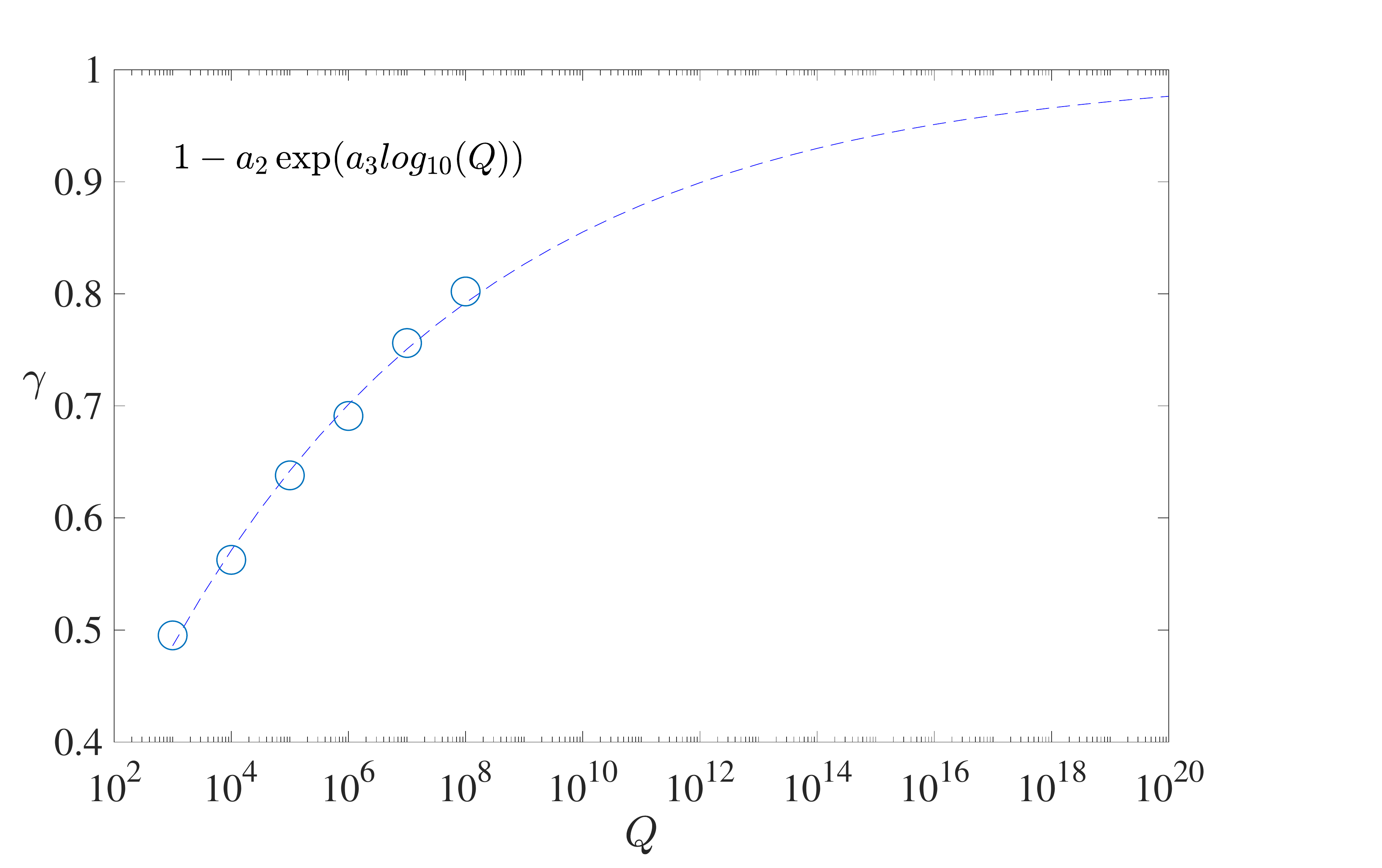} }

 \end{center}

\caption{Power-law fits to the heat transport scaling within the columnar regime for $Pr=1$. (a) The compensated Nusselt number, $Nu/(a_1(Ra/Q)^\gamma)$, versus the Rayleigh number, $Ra$, where 
$\gamma=1-a_2\exp(a_3 \log_{10}(Q))$. (b) $\gamma$ plotted versus $Q$. Here $a_1= 0.4088$, $a_2=0.8847$, and $a_3=-0.1810$. }
\label{F:gamma}
\end{figure*}

\subsection{Power-law fits}

The $Pr=1$ heat transport data shown in Fig.~\ref{F:NuRa}(c) suggests that a power-law scaling of the form $Nu = a_1 (Ra/Q)^{\gamma}$ (where $a_1$ is a constant) is present in the second, columnar regime, where $\gamma$ appears to approach unity for increasing $Q$. 
Since  $\gamma$ increases with $Q$ at a decreasing rate, we first compute $\gamma$ in the columnar regime for individual $Q$ values, then compute a least-squares fit of the form  $\gamma=1-a_2 \exp(a_3\log_{10}(Q))$, where $a_2$ and $a_3$ are constants.  This latter exponential fit for $\gamma$ has no physical basis and is meant only to provide a guide for the behavior at large values of $Q$. Fig.\ref{F:gamma} (a) shows the rescaled, or compensated, Nusselt number, $Nu/(a_1(Ra/Q)^\gamma)$, plotted versus the Rayleigh number, $Ra$, with the fitting results:  $a_1= 0.4088$, $a_2=0.8847$, $a_3=-0.1810$.

Fig.~\ref{F:gamma}(b) shows in detail how $\gamma$ changes with $Q$. The exponential fit suggests that $\gamma$ reaches $0.95 $ around $Q=10^{16}$, which is close to estimates for $Q$ in the Earth's outer core \citep[e.g.][]{nG10}.

\begin{figure*}
\begin{center}
\hspace*{-0.6cm} 
         \subfloat[]{ 
             \includegraphics[height=5.5cm]{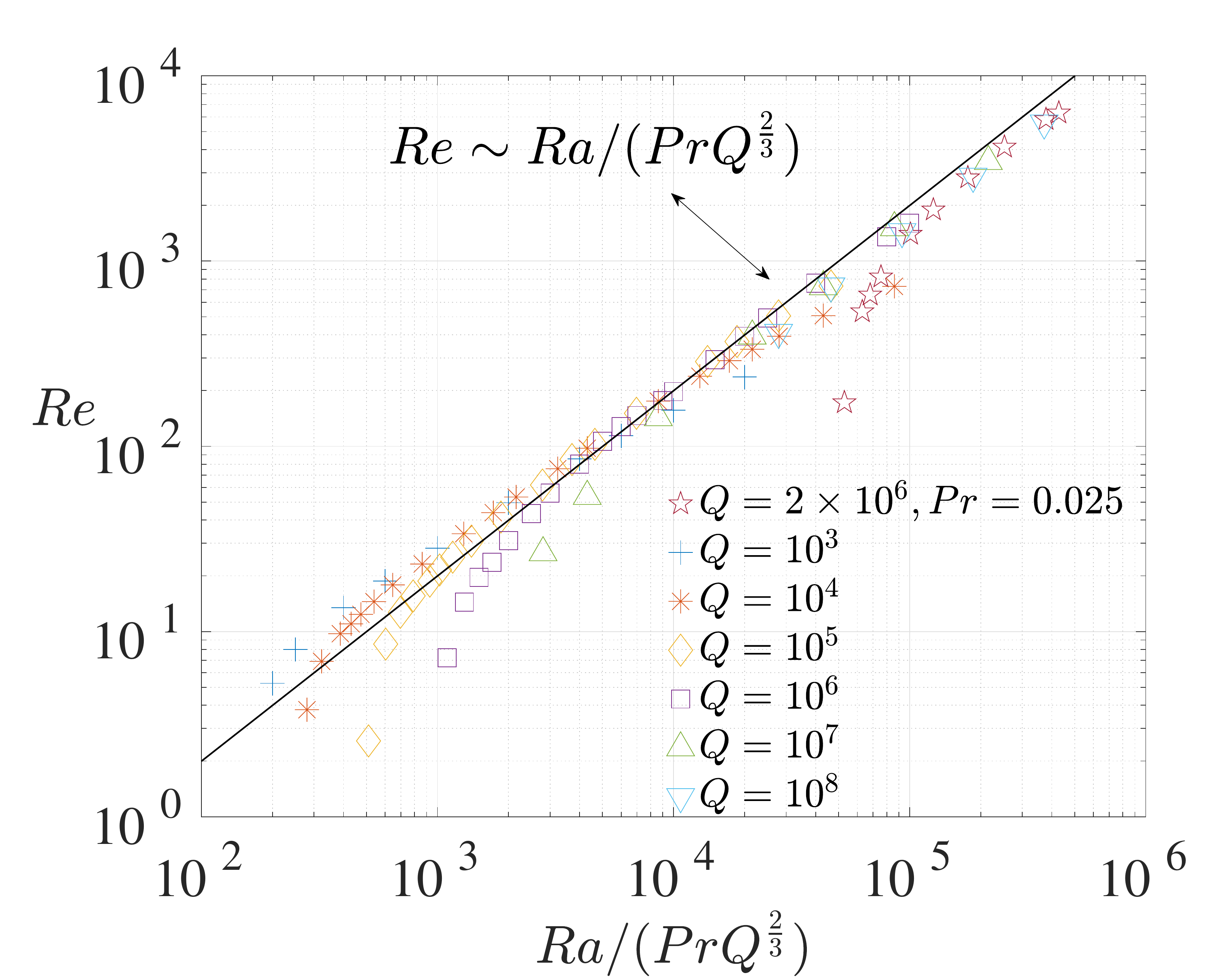}} 
        \subfloat[]{ 
\includegraphics[height=5.5cm]{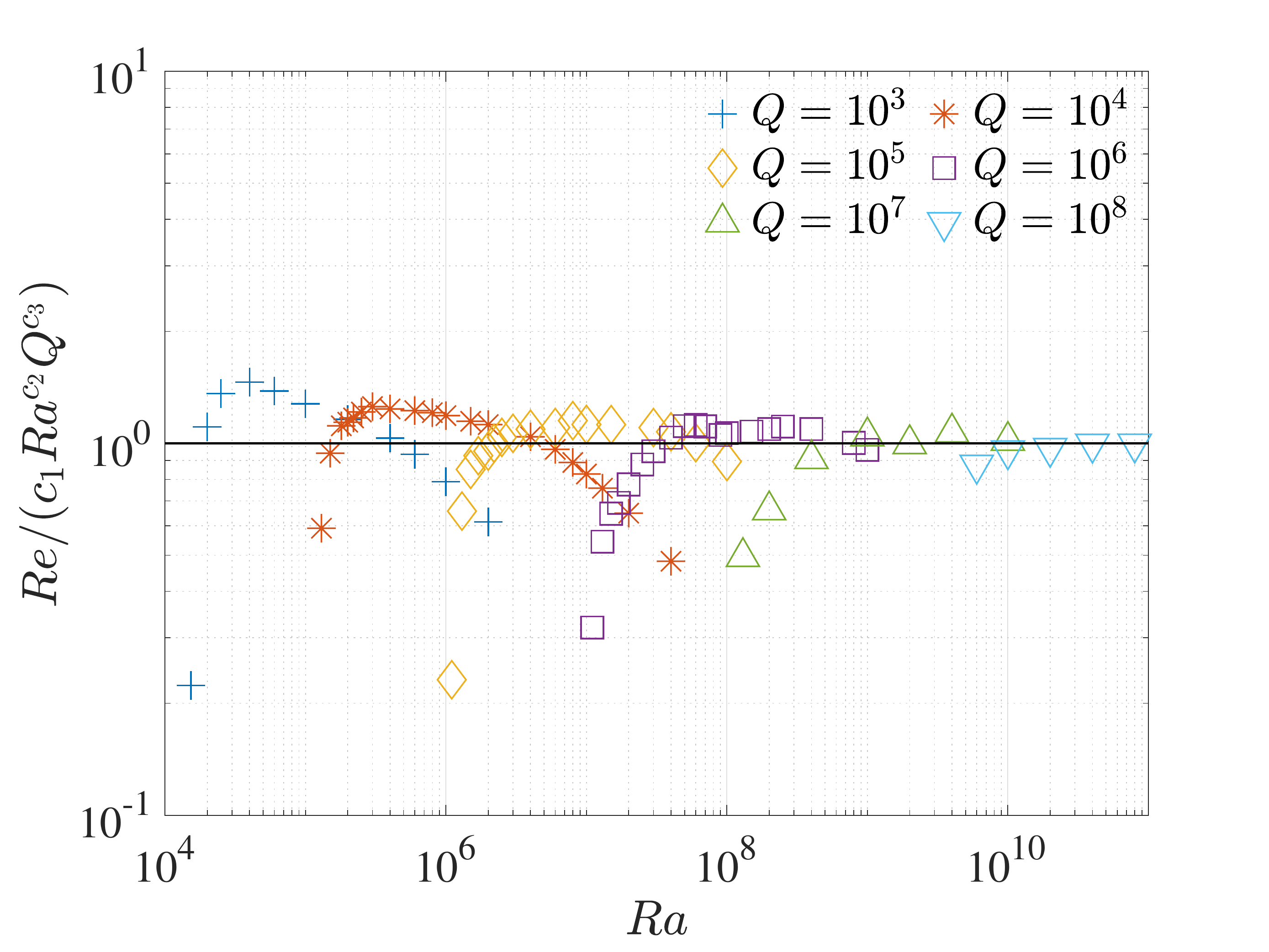} }
 \end{center}
\caption{Power-law scaling for the Reynolds number. (a) The Reynolds number versus $Ra/(PrQ^{2/3})$. (b) The compensated Reynolds number, $Re/(c_1Ra^{c_2}Q^{c_3})$, plotted versus the Rayleigh number, $Ra$. Here $c_1= 0.0324,  c_2= 0.9542, c_3=-0.6461$ are given by a  least squares fitting of $Re=c_1Ra^{c_2}Q^{c_3}$  applied to the $Re$ data in the columnar regime.  }
\label{F:ReRaQ_1}
\end{figure*}

The flow speeds in the columnar regime can be understood by balancing the Lorentz force $F_{l}$ and buoyancy force $F_b$ in the vertical component of the momentum equation, such that
\be   Q \partial_z  b_z \approx (Ra/Pr) T'   \quad \Rightarrow \quad  Q  b_z \sim (Ra/Pr) T', 
\ee
where we again assume $\ell \sim Q^{-1/6}$, vertical derivatives are order one, $u_z \sim Re$ and $ |b_z| \sim  \ell^2 Re $.
If we also assume a weak $Q$-dependence of $T'$ we then have 
\be      Q  \ell^{2} Re \sim Ra/Pr \quad \Rightarrow \quad Re\sim Ra/(PrQ^{2/3}) . 
\ee
We emphasize that this magnetically-constrained scaling is steeper than the turbulent, free-fall scaling in which $Re \sim Ra^{1/2}$ due to the linearity of the balance. In Fig.\ref{F:ReRaQ_1} (a) we show that this scaling collapses the data  well in the columnar regime. A  least squares fitting of $Re=c_1Ra^{c_2}Q^{c_3}$  applied to the $Re$ data in the columnar regime gives $c_1= 0.0324$,  $c_2= 0.9542$, and $c_3=-0.6461$. Fig.\ref{F:ReRaQ_1} (b) shows the compensated Reynolds number, $c_1^{-1}ReRa^{-c_2}Q^{-c_3}$, plotted versus $Ra$. As might be expected if one assumes an asymptotic state exists, the coefficients fit the data better as $Q$ increases.

\section{Conclusions}
A systematic parameter survey of quasi-static magnetoconvection was carried out in a plane layer geometry. The results show three primary magnetoconvection regimes, with each regime distinguishable through unique heat transfer and convective-speed scalings, flow morphology, spectral characteristics, and dominant balances in the momentum and heat equations. For a fixed value of $Q$, and in order of increasing Rayleigh number, the first two regimes are characterized by a predominant Lorentz force and are therefore magnetically-constrained; the third regime is transitional and weakly-influenced by the Lorentz force. For large values of $Q$, the convective flow is highly-anisotropic in both the first (cellular) and second (columnar) regimes. The columnar regime is characterized by spatially-localized convective columns that span the fluid depth, and numerical analysis of the governing equations demonstrates that this regime is characterized by asymptotically-small advection and inertia, despite large Reynolds numbers.

Heat transport in the columnar regime is controlled by the thermal boundary layers and shows a power-law scaling with the Rayleigh number. Previous MC studies have suggested a $Nu \sim (Ra/Q)$ scaling for the $Q \rightarrow \infty$ limit. More generally, our simulations suggest a scaling of the form $Nu \sim (Ra/Q)^{\gamma}$, with $\gamma \rightarrow 1$ as $Q \rightarrow \infty$. Thus, previous work finding $\gamma = 1/2$ \citep{jmA01,xY18} and $\gamma=2/3$ \citep{uB01} are transitional and observable only at relatively small values of $Q$.

For non-magnetic convection, the so-called `ultimate' regime is a hypothetical state of convection in which the entire fluid layer is turbulent, and is thought to arise in the asymptotic limit of large $Ra$ \citep{rK62}. Simulations of unconstrained convection in triply-periodic domains have shown that the ultimate regime may occur in the absence of thermal boundary layers \citep{dL03}, and recent strongly-forced, two-dimensional simulations in a bounded domain also show a transition in heat transport that is indicative of an ultimate regime \citep{xZ18}. Laboratory experiments using internal heating also show evidence of reaching an ultimate regime \citep{sL18}.
When additional forces are present that constrain the convection, regimes of flow that are fundamentally different from those in unconstrained convection can be realized. Our results suggest that the columnar regime represents an ultimate state for magnetically-constrained, quasi-static MC. Of course, the relatively limited accessible values of $Q$ (and $Ra$) might hinder the ability to observe the $Nu \sim Ra/Q$ scaling that is thought to be indicative of this MC state, though our data suggests a trend toward this limit.  

Finally, the Coriolis force, like the Lorentz force, can also act as a `constraint' on convection that results in anisotropic flows. In contrast to MC, however, rotationally-constrained convection does exhibit a turbulent state, as observed with direct numerical simulation \citep[e.g.][]{sS14,cG14,bF14} and numerical simulation of an asymptotically-reduced equation set \citep{kJ12,aR14}. This difference may be due to the different energetic contributions that each of these two forces makes to convection, and to the different asymptotic scalings that characterize the two types of convection. While the Coriolis force has zero direct contribution to the energetics of rotating convection, the quasi-static Lorentz force is purely dissipative. Moreover, all three components of the velocity vector are of comparable magnitude in rotationally-constrained convection \citep{mS06}, yet in MC the vertical component of the velocity vector is asymptotically-larger than the corresponding horizontal components that results in different asymptotic-ordering of the various forces in the momentum equation \citep{pM99}. In particular, in large-$Q$ MC, the viscous force is asymptotically-larger than the inertial force, whereas these two terms are of the same asymptotic order in rotationally-constrained convection. Understanding how the Lorentz and Coriolis forces act in combination on convection is important in understanding magnetic field generation in stars and planets. Although numerical simulations have led to significant advances in our understanding of the role played by these two forces in convection \citep[e.g.][]{rY16,nS17,jA17}, it is currently unknown what ultimate state appears at high Rayleigh numbers, or if such a state exists at all.

\section*{Acknowledgements}
This work was supported by the National Science Foundation under grant EAR \#1620649 (MY, MAC, SM and KJ). SMT was supported by funding from the European Research Council (ERC) under the European Union’s Horizon 2020 research and innovation program (agreement no. D5S-DLV-786780). This work utilized the RMACC Summit supercomputer, which is supported by the National Science Foundation (awards ACI-1532235 and ACI-1532236), the University of Colorado Boulder, and Colorado State University. The Summit supercomputer is a joint effort of the University of Colorado Boulder and Colorado State University. The authors acknowledge the Texas Advanced Computing Center (TACC) at The University of Texas at Austin for providing high-performance computing resources that have contributed to the research results reported within this paper. Volumetric rendering was performed with the visualization software VAPOR \citep{jC05,jC07}.
\newpage

\bibliographystyle{jfm}
\bibliography{References}
\section*{Appendix}
Here we provide tables with details of the numerical simulations.

\newpage

\begin{table}
\begin{center}
\begin{tabular}{c c c c c c c c c c c}
\hline
case & $Q$  &  $Ra$ & $Nu$   &  $Re$ & $\Delta t$ &$N_x \times N_y \times N_z$ \\
\hline 
a1 & $0$  & $1 \times 10^4$ & $3.62\pm 0.03$& $37.46\pm 0.22$&$5\times10^{-4}$ & $384 \times 384 \times 48 $ \\
a2 & $0$  & $2 \times 10^4$ & $4.41\pm 0.04$& $53.49\pm 0.38$ &$2\times10^{-4}$ & $384 \times 384 \times 96 $ \\
a3 & $0$  & $4 \times 10^4$ & $5.37\pm 0.05$& $75.77\pm 0.59$&$5\times10^{-5}$ & $768 \times 768 \times 96 $ \\
a4 & $0$  & $1 \times 10^5$ & $6.93\pm 0.05$& $115.99\pm 0.67$ &$5\times10^{-5}$ & $768 \times 768 \times 96 $ \\
a5 & $0$  & $2 \times 10^5$ & $8.46\pm 0.06$& $159.69\pm 0.52$&$5\times10^{-5}$ & $768 \times 768 \times 96 $ \\
a6 & $0$  & $4 \times 10^5$ &$10.34\pm 0.07$& $220.35\pm 0.86$&$2\times10^{-5}$ & $768 \times 768 \times 96 $ \\
a7* & $0$  & $1 \times 10^6$ &$13.51\pm 0.14$& $333.47\pm 2.84$ &$5\times10^{-6}$ & $768 \times 768 \times 144 $ \\
a8* & $0$  & $2 \times 10^6$ &$16.62\pm 0.13$& $458.52\pm 3.11$&$5\times10^{-7}$ & $1296 \times 1296 \times 144 $ \\
\hline
\end{tabular}
\end{center}
\caption{\textbf{Details of the RBC cases.} $\Delta t$ is the timestep size.  $N_x \times N_y \times N_z$ denotes the spatial resolution. The box ratio is $10\lambda _c \times 10\lambda _c \times 1$ in cases without *, and $5\lambda _c \times 5\lambda _c \times 1$ in cases  with * above. The horizontal wavenumber $k_c=2.2215$.}
\end{table}
%\newpage

\begin{table}
\begin{center}
\begin{tabular}{c c c c c c c c c c c}
case & $Q$  &  $Ra$ & $Nu$  &   $Re$ & $\Delta t$ &$N_x \times N_y \times N_z$ \\
\hline 
b1 & $10^3$  & $1.53 \times 10^4$ & $1.01\pm 0.00^*$ & $0.82\pm 0.00^*$&$2\times10^{-3}$ & $96 \times 96 \times 48 $ \\
b2 & $10^3$  & $2 \times 10^4$ & $1.46\pm 0.00^*$& $5.26\pm 0.00^*$&$2\times10^{-3}$ & $96 \times 96 \times 48 $ \\
b3 & $10^3$  & $2.5 \times 10^4$ & $1.80\pm 0.01$& $8.00\pm 0.10$ &$1\times10^{-3}$ & $144 \times 144 \times 48 $ \\
b4 & $10^3$  & $4 \times 10^4$ & $2.50\pm 0.02$ & $13.45\pm 0.24$ &$5\times10^{-4}$ & $144 \times 144 \times 48 $ \\
b5 & $10^3$  & $6 \times 10^4$ & $3.05\pm 0.02$& $18.74\pm 0.31$&$2\times10^{-4}$ & $192 \times 192 \times 72 $ \\
b6 & $10^3$  & $1 \times 10^5$ & $3.87\pm 0.03$ & $28.18\pm 0.32$ &$1\times10^{-4}$ & $288 \times 288 \times 144 $ \\
b7 & $10^3$  & $2 \times 10^5$ & $5.48\pm 0.05$& $49.60\pm 0.70$&$4\times10^{-5}$ & $288 \times 288 \times 144 $  \\
b8 & $10^3$  & $4 \times 10^5$ & $7.80\pm 0.08$& $85.55\pm 1.02$&$2\times10^{-5}$ & $384 \times 384 \times 144 $ \\
b9 & $10^3$  & $6 \times 10^5$ & $9.45\pm 0.10$& $113.97\pm 1.34$&$2\times10^{-5}$ & $384 \times 384 \times 144 $  \\
b10 & $10^3$  & $1 \times 10^6$ & $11.62\pm 0.13$& $156.48\pm 1.39$&$1\times10^{-5}$ & $576 \times 576 \times 144 $ \\
b11 & $10^3$  & $2 \times 10^6$ & $14.97\pm 0.14$ & $236.63\pm 1.96$&$1\times10^{-5}$ & $576 \times 576 \times 144 $ \\
\hline
\end{tabular}
\end{center}
\caption{\textbf{Details of the $Q=10^3$ cases.} $\Delta t$ is the timestep size.  $N_x \times N_y \times N_z$ denotes the spatial resolution. The box ratio is $10\lambda _c \times 10\lambda _c \times 1$. The horizontal wavenumber $k_c=5.6842$. The case with  $\pm 0.00^*$ indicates that it is just above critical Rayleigh and has a stable $Nu$ or $Re$.}
\end{table}

\begin{table}
\begin{center}
\begin{tabular}{c c c c c c c c c c c}
case & $Q$  &  $Ra$ & $Nu$  &  $Re$ & $\Delta t$ &$N_x \times N_y \times N_z$ \\
\hline 
c1 & $10^4$  & $1.3 \times 10^5$ & $1.149\pm 0.002$ & $3.79\pm 0.02$ &$1\times10^{-4}$ & $96 \times 96 \times 48 $ \\
c2 & $10^4$  & $1.5 \times 10^5$ & $1.43\pm 0.01$& $6.91\pm 0.10$ &$1\times10^{-4}$ & $96 \times 96 \times 48 $ \\
c3 & $10^4$  & $1.8 \times 10^5$ & $1.78\pm 0.01$& $9.74\pm 0.09$ &$1\times10^{-4}$ & $96 \times 96 \times 48 $\\
c4 & $10^4$  & $2 \times 10^5$ & $1.98\pm 0.01$& $11.01\pm 0.20$&$1\times10^{-4}$ & $192 \times 192 \times 48 $ \\
c5 & $10^4$  & $2.2\times 10^5$ & $2.17\pm 0.01$& $12.38\pm 0.24$&$1\times10^{-4}$ & $192 \times 192 \times 48 $\\
c6 & $10^4$  & $2.5 \times 10^5$ & $2.42\pm 0.02$& $14.51\pm 0.28$&$1\times10^{-4}$ & $192 \times 192 \times 48 $ \\
c7 & $10^4$  & $3 \times 10^5$ & $2.76\pm 0.02$& $17.85\pm 0.36$ &$1\times10^{-4}$ & $192 \times 192 \times 48 $ \\
c8 & $10^4$  & $4 \times 10^5$ & $3.30\pm 0.04$& $23.15\pm 0.86$ &$5\times10^{-5}$ & $192 \times 192 \times 48 $ \\
c9 & $10^4$  & $6\times 10^5$ & $4.17\pm 0.05$& $33.74\pm 0.87$&$5\times10^{-5}$ & $192 \times 192 \times 48 $\\
c10 & $10^4$  & $8 \times 10^5$ & $4.97\pm 0.05$& $43.84\pm 1.02$&$5\times10^{-5}$ & $288 \times 288 \times 48 $ \\
c11 & $10^4$  & $1 \times 10^6$ & $5.62\pm 0.04$ & $53.26\pm 0.96$&$5\times10^{-5}$ & $288 \times 288 \times 48 $ \\
c12 & $10^4$  & $1.5 \times 10^6$ & $7.17\pm 0.04$ & $75.72\pm 0.88$ &$2\times10^{-5}$ & $288 \times 288 \times 48 $\\
c13 & $10^4$  & $2 \times 10^6$ & $8.48\pm 0.05$& $97.62\pm 1.35$&$2\times10^{-5}$ & $384 \times 384 \times 48 $ \\
c14 & $10^4$  & $4 \times 10^6$ & $12.66\pm 0.09$ & $175.40\pm 1.89$&$2\times10^{-5}$ & $576 \times 576 \times 72$ \\
c15 & $10^4$  & $6 \times 10^6$ & $15.71\pm 0.10$& $238.97\pm 2.04$ &$5\times10^{-6}$ & $768 \times 768 \times 72 $ \\
c16* & $10^4$  & $8 \times 10^6$ & $18.07\pm 0.23$& $289.83\pm 3.26$&$4\times10^{-6}$ & $576 \times 576 \times 96 $ \\
c17* & $10^4$  & $1 \times 10^7$ & $20.07\pm 0.33$& $333.93\pm 3.48$&$2\times10^{-6}$ & $576 \times 576 \times 144 $ \\
c18* & $10^4$  & $1.3 \times 10^7$ & $22.40\pm 0.33$& $382.81\pm 5.97$&$1\times10^{-6}$ & $576 \times 576 \times 144 $ \\
c19* & $10^4$  & $2 \times 10^7$ & $27.09\pm 0.35$& $507.76\pm 9.77$ &$2\times10^{-7}$ & $768 \times 768 \times 192 $ \\
c20* & $10^4$  & $4 \times 10^7$ & $34.27\pm 0.37$& $730.47\pm 8.33$ &$1\times10^{-7}$ & $1152 \times 1152 \times 288 $ \\

\hline
\end{tabular}
\end{center}
\caption{\textbf{Details of the $Q=10^4$ cases.} $\Delta t$ is the timestep size.  $N_x \times N_y \times N_z$ denotes the spatial resolution. The box ratio is $10\lambda _c \times 10\lambda _c \times 1$ in cases without *, and $5\lambda _c \times 5\lambda _c \times 1$ in cases  with * above. The horizontal wavenumber $k_c=8.6062$.}
\end{table}
\newpage

\begin{table}
\begin{center}
\begin{tabular}{c c c c c c c c c c c}
case & $Q$  &  $Ra$ & $Nu$  &  $Re$& $\Delta t$ &$N_x \times N_y \times N_z$ \\
\hline 
d1 & $10^5$  & $1.1 \times 10^6$ & $1.04\pm 0.00^*$& $2.57\pm 0.02$  &$2\times10^{-5}$ & $96 \times 96 \times 48 $ \\
d2 & $10^5$  & $1.3 \times 10^6$ & $1.384\pm 0.001$ & $8.56\pm 0.05$ &$2\times10^{-5}$ & $96 \times 96 \times 48 $ \\
d3 & $10^5$  & $1.5 \times 10^6$ & $1.717\pm 0.002$& $12.70\pm 0.14$ &$2\times10^{-5}$ & $96 \times 96 \times 48 $ \\
d4 & $10^5$  & $1.7 \times 10^6$ & $2.01\pm 0.01$& $15.58\pm 0.23$&$2\times10^{-5}$ & $96 \times 96 \times 48 $ \\
d5 & $10^5$  & $2 \times 10^6$ & $2.39\pm 0.01$& $18.71\pm 0.39$&$2\times10^{-5}$ & $96 \times 96 \times 48 $ \\
d6 & $10^5$  & $2.2 \times 10^6$ & $2.64\pm 0.02$& $21.51\pm 0.23$ &$2\times10^{-5}$ & $96 \times 96 \times 48 $ \\
d7 & $10^5$  & $2.5 \times 10^6$ & $2.98\pm 0.02$& $25.21\pm 0.51$&$2\times10^{-5}$ & $192 \times 192 \times 48 $ \\
d8 & $10^5$  & $3 \times 10^6$ & $3.45\pm 0.02$& $30.76\pm 0.48$&$2\times10^{-5}$ & $192 \times 192 \times 48 $ \\
d9 & $10^5$  & $4 \times 10^6$ & $4.24\pm 0.04$& $41.51\pm 0.82$&$2\times10^{-5}$ & $192 \times 192 \times 48 $ \\
d10 & $10^5$  & $6\times 10^6$ & $5.55\pm 0.05$& $61.80\pm 1.00$&$1\times10^{-5}$ & $288 \times 288 \times 72 $ \\
d11 & $10^5$  & $8 \times 10^6$ & $6.77\pm 0.05$& $84.85\pm 1.85$&$5\times10^{-6}$ & $384 \times 384 \times 72 $ \\
d12 & $10^5$  & $1 \times 10^7$ & $7.88\pm 0.05$& $102.51\pm 1.47$ &$5\times10^{-6}$ & $384 \times 384 \times 72 $  \\
d13 & $10^5$  & $1.5 \times 10^7$ & $10.33\pm 0.07$ & $150.78\pm 2.33$&$5\times10^{-6}$ & $384 \times 384 \times 72 $  \\
d14 & $10^5$  & $3 \times 10^4$ & $16.42\pm 0.07$& $286.82\pm 3.26$&$5\times10^{-6}$ & $576 \times 576 \times 72 $ \\
d15 & $10^5$  & $4 \times 10^7$ & $19.53\pm 0.08$& $367.83\pm 4.27$ &$5\times10^{-6}$ & $576 \times 576 \times 96 $ \\
d16 & $10^5$  & $6 \times 10^7$ & $24.56\pm 0.11$& $506.18\pm 4.78$ &$2\times10^{-6}$ & $768 \times 768 \times 96 $ \\
d17 & $10^5$  & $1 \times 10^8$ & $32.04\pm 0.12$& $733.46\pm 5.79$ &$5\times10^{-7}$ & $1152 \times 1152 \times 144 $ \\

\hline
\end{tabular}
\end{center}
\caption{\textbf{Details of the $Q=10^5$ cases.} $\Delta t$ is the timestep size.  $N_x \times N_y \times N_z$ denotes the spatial resolution. The box ratio is $10\lambda _c \times 10\lambda _c \times 1$. The horizontal wavenumber $k_c=12.8343$. The case with  $\pm 0.00^*$ indicates that it is just above critical Rayleigh and has a stable $Nu$ or $Re$.}
\end{table}
\newpage

\begin{table}
\begin{center}
\begin{tabular}{c c c c c c c c c c c}
case & $Q$  &  $Ra$ & $Nu$  &$Re$ &  $\Delta t$ &$N_x \times N_y \times N_z$ \\
\hline 
e1 & $10^6$  & $1.1 \times 10^7$ & $1.135\pm 0.001$ & $7.22\pm 0.02$&$2\times10^{-6}$ & $96 \times 96 \times 48 $ \\
e2 & $10^6$  & $1.3 \times 10^7$ & $1.518\pm 0.002$& $14.4\pm 0.10$&$2\times10^{-6}$ & $96 \times 96 \times 48 $ \\
e3 & $10^6$  & $1.5 \times 10^7$ & $1.898\pm 0.004$ & $19.63\pm 0.16$ &$2\times10^{-6}$ & $96 \times 96 \times 48 $ \\
e4 & $10^6$  & $1.7 \times 10^7$ & $2.255\pm 0.004$& $23.64\pm 0.16$ &$2\times10^{-6}$ & $144 \times 144 \times 48 $ \\
e5 & $10^6$  & $2 \times 10^7$ & $2.78\pm 0.004$ & $30.95\pm 0.19$&$2\times10^{-6}$ & $144 \times 144 \times 48 $ \\
e6 & $10^6$  & $2.5 \times 10^7$ & $3.5\pm 0.006$& $43.35\pm 0.20$&$2\times10^{-6}$ & $192 \times 192 \times 48 $ \\
e7 & $10^6$  & $3   \times 10^7$ & $4.28\pm 0.02$& $55.99\pm 0.17$ &$2\times10^{-6}$ & $192 \times 192 \times 48 $ \\
e8 & $10^6$  & $4 \times 10^7$ & $5.43\pm 0.02$& $80.13\pm 0.34$&$2\times10^{-6}$ & $192 \times 192 \times 48 $ \\
e9 & $10^6$  & $5 \times 10^7$ & $6.56\pm 0.02$& $106.47\pm 0.35$&$2\times10^{-6}$ & $288 \times 288 \times 48 $ \\
e10 & $10^6$  & $6 \times 10^7$ & $7.43\pm 0.03$& $127.64\pm 0.34$&$2\times10^{-6}$ & $288 \times 288 \times 48 $ \\
e11 & $10^6$  & $7 \times 10^7$ & $8.25\pm 0.05$ & $146.50\pm 0.85$&$2\times10^{-6}$ & $384 \times 384 \times 48 $ \\
e12 & $10^6$  & $9 \times 10^7$ & $9.74\pm 0.04$& $176.64\pm 1.03$ &$2\times10^{-6}$ & $384 \times 384 \times 48 $ \\
e13 & $10^6$  & $1 \times 10^8$ & $10.43\pm 0.06$& $197.42\pm 1.72$&$2\times10^{-6}$ & $384 \times 384 \times 48 $ \\
e14 & $10^6$  & $1.5 \times 10^8$ & $14.03\pm 0.07$& $293.64\pm 2.35$&$2\times10^{-6}$ & $384 \times 384 \times 48 $ \\
e15 & $10^6$  & $2 \times 10^8$ & $17.36\pm 0.07$& $392.51\pm 3.53$&$2\times10^{-6}$ & $576 \times 576 \times 72 $ \\
e16 & $10^6$  & $2.5 \times 10^8$ & $20.49\pm 0.09$& $493.42\pm 4.08$&$1\times10^{-6}$ & $576 \times 576 \times 96 $ \\
e17 & $10^6$  & $4 \times 10^8$ & $28.58\pm 0.09$& $760.12\pm 7.85$ &$1\times10^{-6}$ & $768 \times 768 \times 96 $ \\
e18 & $10^6$  & $8 \times 10^8$ & $43.83\pm 0.13$& $1.35\times 10^3\pm 8.15$&$5\times10^{-7}$ & $1152 \times 1152 \times 144$ \\
e19 & $10^6$  & $1 \times 10^9$ & $49.70\pm 0.16$& $1.60\times 10^3\pm 7.92$&$2\times10^{-7}$ & $1536 \times 1536 \times 144 $ \\

\hline
\end{tabular}
\end{center}
\caption{\textbf{Details of the ${Q=10^6}$ cases.} $\Delta t$ is the timestep size.  $N_x \times N_y \times N_z$ denotes the spatial resolution. The box ratio is $10\lambda _c \times 10\lambda _c \times 1$. The horizontal wavenumber $k_c=18.9823$.}
\end{table}
%\newpage

\begin{table}
\begin{center}
\begin{tabular}{c c c c c c c c c c c}
case & $Q$  &  $Ra$ & $Nu$  &   $Re$ &$\Delta t$ &$N_x \times N_y \times N_z$ \\
\hline 
f1 & $10^7$  & $1.3 \times 10^8$ & $1.615\pm 0.002$ & $26.65\pm 0.11$ &$2\times10^{-7}$ & $192 \times 192 \times 48 $ \\
f2 & $10^7$  & $2 \times 10^8$ & $3.124\pm 0.01$& $53.66\pm 0.48$ &$2\times10^{-7}$ & $192 \times 192 \times 48 $ \\
f3 & $10^7$  & $4 \times 10^8$ & $6.37\pm 0.03$& $142.80\pm 1.28$ &$2\times10^{-7}$ & $192 \times 192 \times 48 $ \\
f4 & $10^7$  & $1 \times 10^9$ & $13.33\pm 0.05$& $393.95\pm 2.38$&$2\times10^{-7}$ & $384 \times 384 \times 96 $ \\
f5 & $10^7$  & $2 \times 10^9$ & $22.29\pm 0.07$& $727.46\pm 5.54$ &$2\times10^{-7}$ & $576 \times 576 \times 96 $ \\
f6 & $10^7$  & $4 \times 10^9$ & $38.26\pm 0.07$ & $1.51\times 10^3\pm 7.16$&$2\times10^{-7}$ & $768 \times 768 \times 144 $ \\
f7 & $10^7$  & $1 \times 10^{10}$ & $72.40\pm 0.13$ & $3.45\times 10^3\pm 11.77$&$5\times10^{-8}$ & $1536 \times 1536 \times 192 $ \\

\hline
\end{tabular}
\end{center}
\caption{\textbf{Details of the ${Q=10^7}$ cases.} $\Delta t$ is the timestep size.  $N_x \times N_y \times N_z$ denotes the spatial resolution. The box ratio is $10\lambda _c \times 10\lambda _c \times 1$. The horizontal wavenumber $k_c=27.9622$.}
\end{table}
%\newpage

\begin{table}
\begin{center}
\begin{tabular}{c c c c c c c c c c c}
case & $Q$  &  $Ra$ & $Nu$  & $Re$  & $\Delta t$ &$N_x \times N_y \times N_z$ \\
\hline 
g1 & $10^8$  & $6 \times 10^9$ & $10.29\pm 0.07$ & $411.91\pm 4.75$ &$2\times10^{-8}$ & $192 \times 192 \times 72 $ \\
g2 & $10^8$  & $1 \times 10^{10}$ & $15.93\pm 0.09$ & $733.33\pm 5.85$ &$2\times10^{-8}$ & $384 \times 384 \times 96 $ \\
g3 & $10^8$  & $2 \times 10^{10}$ & $27.60\pm 0.17$& $1.44\times 10^3\pm 19.84$ &$2\times10^{-8}$ & $384 \times 384 \times 144 $ \\
g4 & $10^8$  & $4 \times 10^{10}$ & $48.05\pm 0.33$& $2.87\times 10^3\pm 29.24$&$2\times10^{-8}$ & $768 \times 768 \times 144 $ \\
g5 & $10^8$  & $8 \times 10^{10}$ & $82.63\pm 0.37$ & $5.56\times 10^3\pm 50.14$ &$2\times10^{-8}$ & $768 \times 768 \times 192 $ \\
\hline
\end{tabular}
\end{center}
\caption{\textbf{Details of the ${Q=10^8}$ cases.} $\Delta t$ is the timestep size.  $N_x \times N_y \times N_z$ denotes the spatial resolution. The box ratio is $5\lambda _c \times 5\lambda _c \times 1$. The horizontal wavenumber $k_c=41.1115$.}
\end{table}

%\newpage

\begin{table}
\begin{center}
\begin{tabular}{c c c c c c c c c c c}
case & $Q$  &  $Ra$ & $Nu$ &  $Re$ & $\Delta t$ &$N_x \times N_y \times N_z$ \\
\hline 
h1 & $2\times10^6$  & $2.1 \times 10^7$ & $1.048\pm 0.003$ & $172.19\pm 2.15$ &$1\times10^{-6}$ & $96 \times 96 \times 48 $ \\
h2 & $2\times10^6$  & $2.5 \times 10^7$ & $1.386\pm 0.004$ & $533.31\pm 10.77$  &$1\times10^{-6}$ & $192 \times 192 \times 48 $ \\
h3 & $2\times10^6$  & $2.7 \times 10^7$ & $1.56\pm 0.01$  & $656.85\pm 13.08$ &$1\times10^{-6}$ & $192 \times 192 \times 48 $ \\
h4 & $2\times10^6$  & $3 \times 10^7$ & $1.81\pm 0.01$ & $821.53\pm 19.56$ &$1\times10^{-6}$ & $192 \times 192 \times 48 $ \\
h5 & $2\times10^6$  & $4 \times 10^7$ & $2.63\pm 0.02$ & $1.40\times 10^3\pm 24.96$ &$1\times10^{-6}$ & $192 \times 192 \times 48 $ \\
h6 & $2\times10^6$  & $5 \times 10^7$ & $3.38\pm 0.03$ & $1.89\times 10^3\pm 55.04$  &$1\times10^{-6}$ & $192 \times 192 \times 48 $ \\
h7 & $2\times10^6$  & $7 \times 10^7$ & $4.68\pm 0.05$  & $2.82\times 10^3\pm 68.50$  &$5\times10^{-7}$ & $288 \times 288 \times 72 $ \\
h8 & $2\times10^6$  & $1 \times 10^8$ & $6.36\pm 0.06$ & $4.13\times 10^3\pm 72.28$ &$1\times10^{-7}$ & $576 \times 576 \times 144 $ \\
h9* & $2\times10^6$  & $1.5 \times 10^8$ & $8.61\pm 0.08$  & $5.82\times 10^3\pm 85.70$  &$5\times10^{-9}$ & $384 \times 384 \times 144 $ \\
h10* & $2\times10^6$  & $1.7 \times 10^8$ & $9.33\pm 0.12$ & $6.31\times 10^3\pm 128.81$ &$1\times10^{-9}$ & $768 \times 768\times 288$ \\
\hline
\end{tabular}
\end{center}
\caption{\textbf{Details of the ${Pr=0.025, Q=2\times10^6}$ cases.} $\Delta t$ is the timestep size.  $N_x \times N_y \times N_z$ denotes the spatial resolution. The box ratio is $10\lambda _c \times 10\lambda _c \times 1$ in cases without *, and $5\lambda _c \times 5\lambda _c \times 1$ in cases  with * above. The horizontal wavenumber $k_c=18.9823$.}
\end{table}

\end{document}